\documentclass[conference]{IEEEtran}
\usepackage{amsmath, amssymb, amsfonts}
\usepackage{graphicx}
\usepackage{subcaption}
\usepackage{booktabs}
\usepackage{multirow}
\usepackage{url}
\usepackage{comment}
\usepackage{algorithm}
\usepackage{algorithmic}
\usepackage{xcolor}
\usepackage{balance}

\title{Fairness-Sensitive PageRank Approximation}

\author{\IEEEauthorblockN{Mukesh Kumar}
\IEEEauthorblockA{Mehta Family School of \\Data Science and Artificial Intelligence\\
IIT Roorkee, India\\
Email: mukesh\_k@mfs.iitr.ac.in}
\and
\IEEEauthorblockN{Gaurav Dixit}
\IEEEauthorblockA{Mehta Family School of \\Data Science and Artificial Intelligence\\
IIT Roorkee, India\\
Email: gaurav.dixit@mfs.iitr.ac.in}
\and
\IEEEauthorblockN{Akrati Saxena}
\IEEEauthorblockA{LIACS, Leiden University\\
The Netherlands\\
Email: a.saxena@liacs.leidenuniv.nl}}

\begin{document}
\maketitle

\begin{abstract}
Real-world social networks have structural inequalities, including the majority and minorities, and fairness-agnostic centrality measures often amplify these inequalities by disproportionately favoring majority nodes. Fairness-Sensitive PageRank aims to balance algorithmic influence across structurally and demographically diverse groups while preserving the link-based relevance of classical PageRank. However, existing formulations require solving constrained matrix inversions that scale poorly with network size. In this work, we develop an efficient mean-field approximation for Fairness-Sensitive PageRank (FSPR) that enforces group-level fairness through an estimated teleportation (jump) vector, thereby avoiding the costly matrix inversion and iterative optimization. We derive a closed-form approximation of FSPR using the in-degree and group label of nodes, along with the global group proportion. We further analyze intra-class fluctuations by deriving expressions for the variance of approximated FSPR scores. Empirical results on real-world networks demonstrate that the proposed approximation efficiently estimates the FSPR while reducing runtime by an order of magnitude, enabling fairness-constrained ranking at scale.
\end{abstract}


\section{Introduction}
Structural inequalities are deeply embedded in real-world social networks \cite{saxena2024fairsna}. Individuals’ positions in these networks, defined by their connections, community affiliations, and structural roles, shape access to information, visibility, and opportunities \cite{dimaggio2012network}. These inequalities are further reinforced by homophily \cite{saxena2025homophily}, where nodes preferentially connect to others with similar attributes, leading to segregated network structures and unequal exposure across groups. Algorithms operating over such networks often inherit and amplify these pre-existing disparities \cite{stoica2024fairness,saxena2022hm}. This has motivated a rapidly growing body of research in algorithmic fairness, which seeks to ensure that algorithmic outcomes do not reinforce existing biases and produce fair outcomes for everyone \cite{saxena2024fairsna,dong2023fairness}.

One of the foundational tools in network analysis is the PageRank algorithm \cite{brin1998anatomy}, initially developed for web search and later applied to social and information networks to identify influential users \cite{gleich2015pagerank}. PageRank computes a stationary distribution of nodes' importance based on random walks, where each node’s score reflects its probability of being visited in the long run. However, in social networks, PageRank can disproportionately favor nodes in dominant or densely connected communities, thereby reproducing existing inequalities \cite{tsioutsiouliklis2021fairness, espin2022inequality}.

To address this issue, Tsioutsiouliklis et al. \cite{tsioutsiouliklis2021fairness} proposed two fairness-aware PageRank algorithms called Fairness-Sensitive PageRank and Locally Fair PageRank. Fairness-Sensitive PageRank (FSPR) \cite{tsioutsiouliklis2021fairness} modifies the PageRank process to incorporate group-level fairness constraints. The objective is to adjust the teleportation (jump) vector so that the total PageRank mass assigned to different demographic groups satisfies predefined fairness targets. They showed that the resulting scores achieve a fairer distribution of algorithmic visibility across groups. Despite the fairness guarantees in the distribution of PageRank weights across groups, it has high computational complexity for large networks.

In this work, we propose an efficient mean-field approximation for Fairness-Sensitive PageRank that integrates degree-based estimation with group-fairness constraint. Our method partitions the network into degree-based communities—classes defined jointly by in-degree, out-degree, and group membership, and derives a mean-field approximation of the FSPR. By approximating the restart probability vector using community proportions and the in-degree of nodes, we eliminate the need to solve the fairness-constrained optimization problem in FSPR. The proposed approximation method is validated on real-world networks, demonstrating its effectiveness in handling large-scale networks.


\section{Fair PageRank Approximation}
\label{sec:FairPR}
The PageRank of a network is expressed in matrix form as:
\begin{equation*}
    \mathbf{p}^\top = (1 - \nu)\, \mathbf{p}^\top \mathbf{P} + \nu\, \mathbf{v}^\top
\end{equation*}  
where $P$ is the transition probability matrix (i.e., the row-normalized adjacency matrix of the network), and $P[i,j]$ denotes the transition probability from node $i$ to node $j$. Sink nodes (nodes without outgoing links) are handled by replacing the corresponding zero rows in $P$ with a uniform probability distribution over all nodes. The parameters $\nu$ and $1-\nu$ denote the teleportation (jump) probability and the probability of following an outgoing link, respectively. During teleportation, the random walker jumps to a node selected according to the jump vector $v$, which defines a probability distribution over all nodes. The teleportation probability is often set to $\nu = 0.15$ \cite{brin1998anatomy}, and the jump vector is set to the uniform distribution. 
 
Tsioutsiouliklis et al. \cite{tsioutsiouliklis2021fairness} defined Fairness-Sensitive PageRank, where fairness is achieved by modifying the teleportation vector $v$ to fulfill the fairness criterion while keeping the transition matrix $P$ fixed. After learning the jump vector for fairness using convex optimization to minimize the utility loss \cite{tsioutsiouliklis2021fairness}, the closed (matrix) form of fair PageRank can be written as the following iterative update equation:

\begin{equation}
p(i) = \nu v_i + (1 - \nu) \sum_{j \to i} \frac{p_{ji}p(j)}{k_{out}(j)}, \quad \ i = 1, \ldots, N
\label{eq:label1}
\end{equation}
where $N$ is the total number of nodes, $j \to i$ indicates a directed link from node $j$ to node $i$, and $k_{\text{out}}(j)$ denotes the out-degree of node $j$. The fair PageRank of a node $i$, denoted as $p(i)$, represents the stationary probability that the random walker visits node $i$, ensured by the nonzero jump probability $\nu$ and the jump vector $v$ that is optimized to achieve fairness in the PageRank algorithm. 

At time step $n$, the probability that a walker is located at a node $i$ is given by the following Markovian equation. 

\begin{equation}
\begin{aligned}
p_n(i) &= \nu v_i + (1 - \nu) \sum_{\substack{j: \, k_{\text{out}}(j) \ne 0}} \frac{a_{ji}}{k_{\text{out}}(j)} \cdot p_{n-1}(j) \\
&\quad + \sum_{\substack{j: \, k_{\text{out}}(j) = 0}} p_{n-1}(j), \quad i = 1, \ldots, N
\end{aligned}
\label{eq:simple1}
\end{equation}
where $a_{ji}$ is the $(j, i)$-th entry of the adjacency matrix, and it is equal to 1 if an edge exists from node $j$ to node $i$, otherwise $0$. The first term in Eq.~(\ref{eq:simple1}) represents teleportation to randomly selected nodes, the second term captures standard random-walk transitions, and the last term accounts for walkers at dangling nodes.

As $n \to \infty$, the contribution of the last term converges to a constant value that affects all nodes equally. Consequently, this term can be eliminated from Eq.~(\ref{eq:simple1}) as long as the resulting scores are appropriately normalized. The stationary solution of Eq.~(\ref{eq:simple1}) defines the Fairness-Sensitive PageRank of node i under the specified fairness constraints, expressed as $p(i) = \lim_{n \to \infty} p_n(i)$.   

\subsection{Mean-Field Approximation}
\label{subsec:mean-field}
For simplicity, we consider networks composed of two groups, $(\text{i})$ protected ($P$), defined by sensitive attributes such as race, gender, and religion, and $(\text{ii})$ unprotected ($U$); as also considered in the FSPR \cite{tsioutsiouliklis2021fairness}.   However, our method can be extended to networks with more than two groups.  

Let's assume that the network is divided into two disjoint groups based on attribute values (protected and unprotected), and approximation is performed independently for each group. Instead of estimating the FSPR of individual nodes, nodes are aggregated into degree-based classes sharing the same in-degree, out-degree, and group membership. 
Specifically, nodes belonging to the protected group are partitioned into degree-classes
\begin{equation*}
\mathcal{K}^{P} =  \{\, k^{P}_{i} \mid i = 1,2,\dots,n' \,\}, 
\end{equation*}
and nodes in the unprotected group are partitioned into
\begin{equation*}
\mathcal{K}^{U} = \{\, k^{U}_{i} \mid i = 1,2,\dots,m' \,\}
\end{equation*}
Each protected degree-class is characterized by a pair $k^{P}_{i} = (k^{P}_{i,\text{in}},\, k^{P}_{i,\text{out}})$, 
and similarly, each unprotected degree-class is defined as $k^{U}_{j} = (k^{U}_{j,\text{in}},\, k^{U}_{j,\text{out}})$, 
where, $k^{P}_{i,\text{in}}$ and $k^{P}_{i,\text{out}}$ denote the in-degree and out-degree of the $i\text{-th}$ degree-class in the protected group, while  $k^{U}_{j,\text{in}}$ and $k^{U}_{j,\text{out}}$ denote the in-degree and out-degree of the $j\text{-th}$ degree-class in the unprotected group. Thus, all nodes within a given degree-class share identical in-degree, out-degree, and group labels.

After aggregating nodes into degree-classes for both groups, the average FSPR of nodes within each class is defined as:

\begin{equation}
    \bar{p}_n(k_{i}^{P}) = \frac{1}{NP(k_{i}^{P})} \sum_{t \in k_{i}^{P}} p_n(t)
    \label{eq:equation4_1}
\end{equation}

\begin{equation}
    \bar{p}_n(k_j^{U}) = \frac{1}{NP(k_{j}^{U})} \sum_{l \in k_{j}^{U}} p_n(l)
    \label{eq:equation4_2}
\end{equation}
where $k_{i}^{P}$ and $k_{j}^{U}$ denote degree-classes of nodes sharing the same in-degree, out-degree, and group label. $N$ is the total count of nodes, while $P(k_{i}^{P} )$ and $P(k_{j}^{U} )$ denote the probabilities that a randomly selected node belongs to the degree-class $k^{P}_{i}$ and $k^{U}_{j}$, respectively. 

Taking an average of Eq.~(\ref{eq:simple1}). over all nodes in degree-class \(k_{i}^{C}\) ($C \in \{P,U\}$), we get: 
\begin{equation}
\frac{1}{NP(k_{i}^{C})} \sum_{t \in k_{i}^{C}} p_n(t) = \nu v_i + \frac{1 - \nu}{NP(k_{i}^{C})} \sum_{t \in k_{i}^{C}} \sum_{\substack{q: \, k_{\text{out}}(q) \ne 0}} \frac{a_{qt}}{k_{\text{out}}(q)} \ p_{n-1}(q)
\label{eq:protected}
\end{equation}

Initially, the modified jump vector components are approximated using the in-degree and the fraction of nodes in each group. Nodes within the same degree-class share identical structural attributes, including in-degree, out-degree, and group label, resulting in identical approximations of the jump vector.
 
Let $\mathcal{K}^{P} = \{k_{1}^{P}, k_{2}^{P}, \ldots, k_{n^{'}}^{P}\}$ and  $\mathcal{K}^{U} = \{k_{1}^{U}, k_{2}^{U}, \ldots, k_{m^{'}}^{U}\}$ denote the degree classes for protected and unprotected nodes, respectively. The fraction of nodes labeled as protected and unprotected is represented as  $\phi$ and $1 - \phi$, where $\phi \in [0, 1]$. 
For a node $v_i$ in degree-class $i$ of the protected group (in-degree $k^{P}_{i,\text{in}}$) and a node $v_j$ in degree-class $j$ of the unprotected group (in-degree $k^{U}_{j,\text{in}}$), the corresponding jump vector components are approximated as:

\begin{equation*}
v_i^{P} = \phi \frac{k^{P}_{i,\text{in}}}{D_P} , 
\qquad
v_j^{U} = (1-\phi)\,\frac{k^{U}_{i,\text{in}}}{D_U} 
\end{equation*}
where $D_{P}$ and $D_{U}$ represent the total in-degree sums of all nodes in the protected and unprotected groups, respectively.

From Eq.~\eqref{eq:equation4_1} and~\eqref{eq:equation4_2}, the left-hand side of Eq ~\eqref{eq:protected} corresponds to the average FSPR \( \bar{p}_n(k_{i}^{C}) \) of nodes within the class $k^{C}_{i}$. On the right side of Eq.~\eqref{eq:protected}, the sum over $q$ is decomposed into two sums: one taken over all degree-classes $k^{'}$, and the other over all nodes within each degree-class $k$. Then, we obtain 
\begin{equation}
\bar{p}_n(k^{C}_{i}) = \nu v^{C}_{i} + \frac{(1 - \nu)}{NP(k^{C}_{i})} 
\sum_{k'} \frac{1}{k'_{\text{out}}} 
\sum_{t \in k^{C}_{i}} 
\sum_{q \in k'} 
a_{qt} \, p_{n-1}(q)
\label{eq:equation7}
\end{equation}

At this stage, we apply the mean-field approximation, where FSPR values of a node's predecessor neighbors are substituted by their degree-class average value. 
\begin{equation}
\sum_{t \in k^{C}_{i}} \sum_{q \in k'} a_{qt} \, p_{n-1}(q)
\simeq
\bar{p}_{n-1}(k') \sum_{t \in k^{C}_{i}} \sum_{q \in k'} a_{qt}
= \bar{p}_{n-1}(k') \, E_{k' \to k^{C}_{i}}
\label{eq:equation6}
\end{equation}
where $E_{k' \to k^{C}_{i}}$ denotes the number of directed edges that originate from nodes associated with degree-class $k'$ and connect to nodes with degree-class $k^{C}_{i}$. This matrix is equivalently expressed as:
\begin{equation}
E_{k' \to k^{C}_{i}} = \frac{k_{i,\text{in}}^{C}  P(k_{i}^{C})  N  E_{k' \to k^{C}_{i}}}{k^{C}_{i,\text{in}}  P(k^{C}_{i})  N}
= k^{C}_{i,\text{in}}  P(k^{C}_{i})  N  P_{\text{in}}(k' \mid k^{C}_{i})
\label{eq:equation9}
\end{equation}
where $P_{\text{in}}(k' \mid k^{C}_{i})$ denotes the conditional probability that a node in degree-class $k^{C}_{i}$ receives a connection from a predecessor node belonging to degree-class $k'$.
 
By substituting Equations ~(\ref{eq:equation6}) and ~(\ref{eq:equation9}) into Eq.~(\ref{eq:equation7}), we obtain
\begin{equation}
\overline{p}_{n}(k^{C}_{i}) = \nu v^{C}_{i} + (1 - \nu) \, k^{C}_{i,\text{in}} \sum_{k'} \frac{P_{\text{in}}(k' \mid k^{C}_{i})}{k'_{\text{out}}(j)}  \overline{p}_{n-1}(k^{'})
\label{eq:equationB13}
\end{equation}

This is a closed-form solution for the average Fairness-Sensitive PageRank of nodes within the same degree-class.
 
In networks without degree-degree correlations (uncorrelated) \cite{pastor2004evolution}, the conditional probability of transitions $P_{\text{in}} \left( k' \,\middle|\, k^{C}_{i} \right)$ becomes independent of degree $k^{C}_{i}$ and reduces to a simpler form, as 
\begin{equation}
P_{\text{in}} \left( k' \,\middle|\, k^{C}_{i} \right) = \frac{k_{\text{out}}'  P(k')}{\langle k_{\text{in}} \rangle}
\label{eq:equation9_1}
\end{equation}
where $\langle k_{\text{in}}\rangle$ is the average in-degree of the network. 

By substituting Eq.~(\ref{eq:equation9}) into Eq.~(\ref{eq:equationB13}) and taking $n \to \infty$, we get:
\begin{equation}
\bar{p}(k^{C}_{i}) = \nu v^{C}_{i} + \frac{(1 - \nu)}{N}  \frac{k^{C}_{i,\text{in}}}{\langle k_{\text{in}} \rangle}
\label{eq:equationB15}
\end{equation}
Hence, the average FSPR of nodes of degree class $k^{T}_{i}$ is independent of $k^{C}_{i,\text{out}}$. 
Here, $C \in \{P, U\}$ denotes the protected and unprotected groups, with $v^{C}_{i} = \phi_{C}\frac{k^{C}_{i,\text{in}}}{D_{C}}$. 
For protected group: $\phi_{C}=\phi$, $D_{C}=D_{P}$, and for unprotected group: $\phi_{C}=1-\phi$, $D_{C}=D_{U}$.

Finally, the approximated FSPR of node $u$ is defined as:   
\begin{equation*}
    \hat{p}(u) = \nu \phi_{C(u)}\frac{k_{\text{in}}(u)}{D_{C(u)}} + (1-\nu) \frac{k_{\text{in}}(u)}{\langle k_{\text{in}}\rangle} 
\end{equation*}
where $\phi_{C(u)}$ is the fraction of nodes in the same group as node $u$, $D_{C(u)}$ is the total in-degree of all nodes within that group, and $k_{\text{in}(u)}$ is the in-degree of node $u$.

\subsection{Fluctuation Analysis}
\label{subsec:Fluctuation}
As derived in Subsection \ref{subsec:mean-field}, the mean-field approximation provides a closed-form expression for the average FSPR of nodes within each degree class. 
However, it does not characterize the fluctuations or distribution of FSPR scores within a given degree class.
To address this, we extend the mean-field approximation to capture fluctuations in fair PageRank values within each degree class. Specifically, we derive an expression for the variance associated with a randomly selected degree class $k^{T}_{i}$. We begin by squaring Eq.~(\ref{eq:simple1}) and omitting the third term as previously discussed in Section \ref{sec:FairPR}, since it represents the contribution of random walkers that were located at dangling nodes in the previous step and subsequently teleported to randomly chosen nodes. In the limit, $n \to \infty$, this term induces a constant effect on all nodes.

\begin{equation}
\begin{aligned}
   p_{n}^{2}(i) = \nu^{2} v_{i}^{2} + 2\nu v_{i}(1-\nu)\sum_{l} \frac{a_{li}}{k_{out}(l)}p_{n-1}(l) + \\ (1-\nu)^{2}\sum_{l} \frac{a_{li}}{k_{out}^{2}(l)}p_{n-1}^{2}(l) + (1-\nu)^{2} \\ \sum_{l \neq l^{'}}\frac{a_{li}a_{l^{'}i}}{k_{out}(l)k_{\mathrm{out}}(l^{'})}p_{n-1}(l)p_{n-1}(l^{'})
\end{aligned}
\label{eq:equationB16}
\end{equation}
As in Eq.(~\ref{eq:equation4_1}), we compute the average of the squared FSPR values over the degree classes, which is defined as follows:
\begin{equation}
    \overline{p^{2}}_{n}(k^{C}_{i}) \equiv  \frac{1}{NP(k^{C}_{i})} \sum_{i \in k^{C}_{i}}p_{n}^{2}(i)
\end{equation}

Averaging Eq.~(\ref{eq:equationB16}) over degree classes and algebraically rearranging the terms, we obtain:
\begin{equation}
\begin{aligned}
    \overline{p^{2}}_{n}(k^{C}_{i}) = \nu^{2} (v^{C}_{i})^{2} + 2\nu v_{i} (1-\nu)k^{C}_{i,\text{in}}\sum_{k^{'}}\frac{P_{\text{in}}(k^{'} \mid k)}{k_{\text{out}}^{'}} + (1-\nu)^{2} \\ \sum_{k^{'}}\frac{P_{\text{in}}(k^{'} \mid k^{C}_{i})}{k_{\text{out}}^{'2}}\overline{p^{2}}_{n}(k^{'}) +  (1-\nu)^{2}k^{C}_{i,\text{in}}(k^{C}_{i,\text{in}}-1) \\ \sum_{k^{'}}\sum_{k^{''}}\frac{P_{\text{in}}(k^{'},k^{''} \mid k^{C}_{i})}{k_{\text{out}}^{'}k_{\text{out}}^{''}}\overline{p}_{n-1}(k^{'}) \overline{p}_{n-1}(k^{''})
\end{aligned}
\end{equation}
Here, we apply the mean-field approximation, as described in Subsection \ref{subsec:mean-field}. The term  $P_{\text{in}}(k^{'},k^{''} \mid k^{C}_{i})$  denotes the joint likelihood that a node in degree-class 
$k^{C}_{i}$ has one incoming link from a node of degree-class $k^{'}$ and another from degree-class $k^{''}$, where $k^{'}$ and $k^{''}$ are degree classes that may belong to either the protected or the unprotected group.
Since the network is assumed to be degree-degree uncorrelated under mean-field approximation, the joint conditional probability decomposes as $P_{\text{in}}(k^{'},k^{''} \mid k^{C}_{i}) = P_{\text{in}}(k^{'} \mid k^{C})P_{\text{in}}(k^{''} \mid k^{C}_{i})$.
Under this assumption, conditioning on the degree-class $k^{C}_{i}$ implies that the probability of having an incoming link from a degree node $k^{''}$ does not influence the presence of an incoming link from a degree node $k^{'}$. 

In this context, we derive an analytical expression for the variance of FSPR values within a degree class, defined as $\sigma_{n}^{2}(k^{C}_{i}) = \overline{p^{2}}_{n}(k^{C}_{i}) - \overline{p}_{n}^{2}(k^{C}_{i})$, as follows:
\begin{equation}
\begin{aligned}
    \frac{\sigma_{n}^{2}(k^{C}_{i})}{(1-\nu)^{2}} = k^{C}_{i,\text{in}}\sum_{k^{'}}\frac{P_{in}(k^{'} \mid k^{C}_{i})}{k_{\text{out}}^{'2}}\sigma_{n-1}^{2}(k^{'}) \\ + k^{C}_{i,\text{in}}\sum_{k^{'}}\frac{P_{\text{in}}(k^{'} \mid k)}{k_{\text{out}}^{'2}}\overline{p}_{n-1}^{2}(k^{'}) \\ - k^{C}_{i,\text{in}}\left[\sum_{k^{'}}\frac{P_{\text{in}}(k^{'} \mid k^{C}_{i})}{k_{\text{out}}^{'}}\overline{p}_{n-1}(k^{'})\right]^{2}
\end{aligned}
\end{equation}
As $n \to \infty$, this equation admits a closed-form solution consistent with the degree-degree uncorrelated assumption.
\begin{equation}
    \sigma^{2}(k^{C}_{i}) = \frac{(1-\nu)^{2}}{N^{2}\langle k_{\text{in}} \rangle^{2}} \frac{\frac{1}{\langle k_{\text{in}} \rangle}\left\langle \frac{(\nu v^{C}_{i}N\langle k_{\text{in}} \rangle) + (1-\nu)k^{C}_{i,\text{in}})^{2}}{k^{C}_{i,\text{out}}} \right\rangle-1}{1 - \frac{(1-\nu)^{2}}{\langle k_{\text{in}} \rangle}\left\langle \frac{k^{C}_{i,\text{in}}}{k^{C}_{i,\text{out}}} \right\rangle}
\end{equation}
For networks with a heavy-tailed in-degree distribution and a large average in-degree, the expression simplifies to:
\begin{equation}
    \sigma^{2}(k^{C}_{i}) \simeq \frac{(1-\nu)^{4}}{N^{2}\langle k_{\text{in}} \rangle^{3}} \left\langle \frac{(k^{C}_{i,\text{in}})^{2}}{k^{C}_{i,\text{out}}} \right\rangle k^{C}_{i,\text{in}}
    \label{eq:equationB23}
\end{equation}
For large in-degree, the coefficient of variation is: 
\begin{equation}
    \frac{\sigma(k^{C}_{i})}{\overline{p}(k^{C}_{i})} = (1-\nu)\left[\left \langle \frac{(k^{C}_{i,\text{in}})^{2}}{k^{C}_{i,\text{out}}} \right\rangle\frac{1}{\langle k_{\text{in}} \rangle k^{C}_{i,\text{in}}} \right]^{1/2}
\label{eq:equationB22}
\end{equation}
In networks with heavy-tailed degree distribution, the factor $\left\langle \frac{(k^{C}_{i,\text{in}})^{2}}{k^{C}_{i,\text{out}}} \right\rangle$ can be large, which indicates that the relative fluctuations are large for small in-degree. In contrast, for large in-degrees, the relative fluctuation becomes less important due to the presence of the term $k^{C}_{i,\text{in}}$ in the denominator. 
The mean-field approximation for the protected and unprotected groups derived in Eq.~(\ref{eq:equationB15}), together with the analogous approximation expression for the unprotected, provides estimates that closely matche the exact FSPR scores. Note that the formulation in Eq.~(\ref{eq:equationB22}) addresses intra-class fluctuations, rather than fluctuations across the entire graph.

\section{Complexity Analysis of FSPR and Mean-Field Approximation}
\label{sec:fspr_MFA_complexity}

In this section, we analyze the time (computational) and space (memory) complexities of the FSPR method and compare them with those of the proposed mean-field approximation. This analysis highlights the scalability limitations of the original FSPR method and demonstrates that the mean-field approximation substantially reduces computational and memory costs while preserving group-level fairness constraints.

\subsection{Complexity of FSPR}
For a directed graph $G=(V, E)$, where $V$ is the set of nodes with cardinality $\lvert V \rvert = N$ and $E$ is the set of edges with cardinality $\lvert E\rvert = M$. 
The FSPR method computes node scores by incorporating a fairness constraint into the closed-form PageRank formulation, resulting in a constraint (quadratic convex) optimization problem. The original FSPR computation consists of two stages: (i) closed-form PageRank computation and (ii) fairness constraints optimization. We analyze the contribution of each stage to the overall complexity.

\textbf{Closed-Form PageRank Computation:} The FSPR first computes the PageRank operator as
\[
Q = \nu \left[ I - (1 - \nu) P \right]^{-1}
\]
where $P \in \mathbb{R}^{N \times N}$ is the transition probability matrix and $\nu$ is the jump factor. Computing the matrix $Q$ explicitly requires inverting an $N \times N$ matrix. In general, matrix inverse incurs a computational complexity of $\mathcal{O}(N^3)$. Although, the transition matrix $P$ is sparse, its matrix $Q$ is dense, introducing dependencies among all nodes and eliminating sparsity. Therefore, the storage of the dense matrix $Q$ and intermediate computation states leads to a quadratic memory complexity of $\mathcal{O}(N^2)$.

\textbf{Fairness-Constrained Optimization:} Given the closed-form PageRank, in the second step, the original FSPR method solves a quadratic convex-constrained optimization problem to compute a jump vector $v$ that satisfies the fairness constraint while minimizing its deviation from the original PageRank scores. The optimization problem is defined over $N(=\lvert V\rvert)$ variables and involves dense matrix operations due to the dependence on $Q$. The objective function is quadratic in the components of the jump vector, and its Hessian matrix is defined by $H(=Q^TQ)$. Since the PageRank operator $Q$ is dense, the resulting Hessian matrix $H$ is also dense. 

Consequently, the quadratic optimization problem is globally coupled among all $N$ components of the jump vector. The dominant computational complexity in this process arises from dense matrix multiplication required to construct $Q^TQ$, which incurs $\mathcal{O}(N^3)$ time complexity. Moreover, storing the Hessian matrix $H$ and intermediate optimization variables requires $\mathcal{O}(N^2)$ memory.

Hence, the overall computational complexity of the FSPR method is $\mathcal{O}(N^3)$ time and $\mathcal{O}(N^2)$ memory, as both the closed-form PageRank computation and fairness-constraints optimization stage demonstrate these scaling behaviors.

\subsection{Complexity of Mean-Field Approximation}
\label{sec:Mean_field_Complexity}
The proposed mean-field approximation avoids explicit matrix inversion and fairness-constrained optimization by expressing FSPR scores as a closed-form function of node degrees and the group's node fraction. In particular, the approximation depends only on the in-degree of nodes, their group membership, and global group-level degree aggregates. Computing these quantities requires a single pass over the edge set to obtain node in-degrees, followed by linear-time aggregation over the nodes. 

Consequently, the overall computational complexity of the mean-field method is $\mathcal{O}(M)$, while the memory requirement is $\mathcal{O}(N + M)$, which corresponds to storing the graph structure together with a small set of auxiliary vectors: the in-degree of each node, a single group label per node, and the resulting PageRank score for each node. Thus linear computation and storage complexity make this approach efficient for large-scale networks with millions of nodes.

Hence, the original FSPR computation has cubic time complexity and quadratic memory requirements due to dense matrix inversion and globally coupled fairness-constrained optimization, making it impractical for large-scale networks. In contrast, the proposed mean-field approximation eliminates these computational bottlenecks by relying solely on node-level degree statistics and group-level information, thereby achieving linear time and space complexity. This reduction from cubic to linear scaling enables fast FSPR computation on large-scale graphs, establishing the mean-field approximation as a significantly more efficient and scalable alternative to the original FSPR.

\section{Experimental Results}

To validate the theoretical results, we conduct extensive experiments on real-world networks to assess how closely the proposed mean-field (MF) approximation matches the exact FSPR scores. Specifically, we evaluate approximation accuracy, the dependence of FSPR on node degree, score fluctuations, and the preservation of group-level fairness constraints.

The real-world datasets are summarized in Table~\ref{tab:table1}, including the total number of nodes and edges, the average in-degree $\langle k_{in} \rangle$, and the fraction of nodes in each group ($r = \phi$ and $b = 1 - \phi$). Group labels are defined differently across datasets. 

The DBLP-Aminer dataset is an undirected coauthorship network constructed from the DBLP bibliography by the Arnetminer academic search system, where nodes correspond to authors and edges indicate coauthored publications. The DBLP-GEN dataset is an undirected coauthorship network derived from DBLP, containing a subset of publications from data mining and database conferences between 2011 and 2020. Twitter is a directed retweet network capturing politically oriented user interactions. The LinkedIn dataset represents a directed social network of professional relationships among users on the LinkedIn platform. The Pokec dataset is a directed social network of users from the Slovak social media platform Pokec, where edges represent friendship links between users. The Web-Google dataset is a directed web graph released by Google in 2002, where each node corresponds to a web page and directed edges represent hyperlinks between them.

For DBLP-Aminer and DBLP-GEN, protected attributes are inferred from author gender using the \texttt{gender-guesser} \cite{santamaria2018comparison} Python package based on first name. Twitter users are labeled by political affiliation following prior work \cite{tsioutsiouliklis2021fairness}. LinkedIn uses gender inferred from public profiles, and Pokec uses self-declared profile attributes. As Web-Google lacks labels, we infer binary groups using the Louvain community detection algorithm \cite{blondel2008fast}, assigning label 1 to the largest community and label 0 to the rest. In our experiments, the damping factor is set to $\nu = 0.15$, following previous works \cite{brin1998anatomy, tsioutsiouliklis2021fairness}.

\begin{table}[t]
  \caption{Summary of real-world datasets}
  \label{tab:table1}
  \centering
\begin{tabular}{l c c c c c}
\hline
Dataset & \#Nodes & \#Edges & $\langle k_{\mathrm{in}} \rangle$ & r & b \\
\hline
Dblp-Gen \cite{tsioutsiouliklis2022link}     & 16501       & 133226    & 8.07  & 0.26 & 0.74 \\
Twitter \cite{tsioutsiouliklis2021fairness}      & 18740      & 61157     & 2.61  & 0.39 & 0.61 \\
Dblp-Aminer \cite{tsioutsiouliklis2020fairness}  & 432469     & 2462422    & 5.81  & 0.18 & 0.82 \\
Web-Google \cite{leskovec2009community}   & 875713     & 5105039    & 5.82  & 0.02 & 0.98  \\
Pokec \cite{takac2012data}        & 1632803    & 30622564   & 18.75 & 0.49 & 0.51 \\
Linkedin  \cite{tsioutsiouliklis2020fairness}    & 3209448    & 13016456   & 4.05  & 0.37 & 0.63  \\
\hline
\end{tabular}
\end{table}

\begin{figure}
    \centering
\begin{subfigure}{0.23\textwidth}
    \centering
    \includegraphics[width=\linewidth]{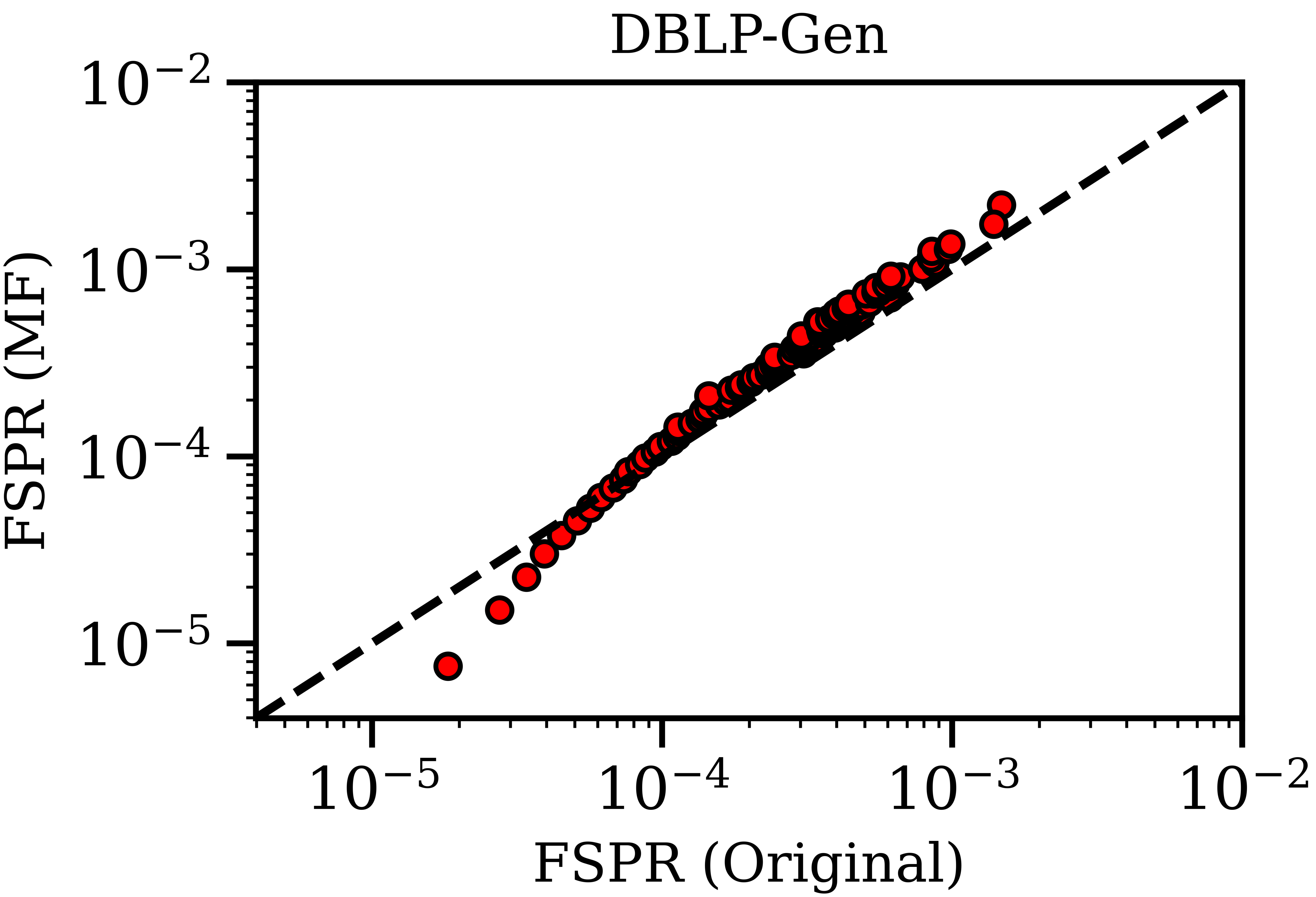}
\end{subfigure}
\vspace{2mm}
\begin{subfigure}{0.23\textwidth}
    \centering
    \includegraphics[width=\linewidth]{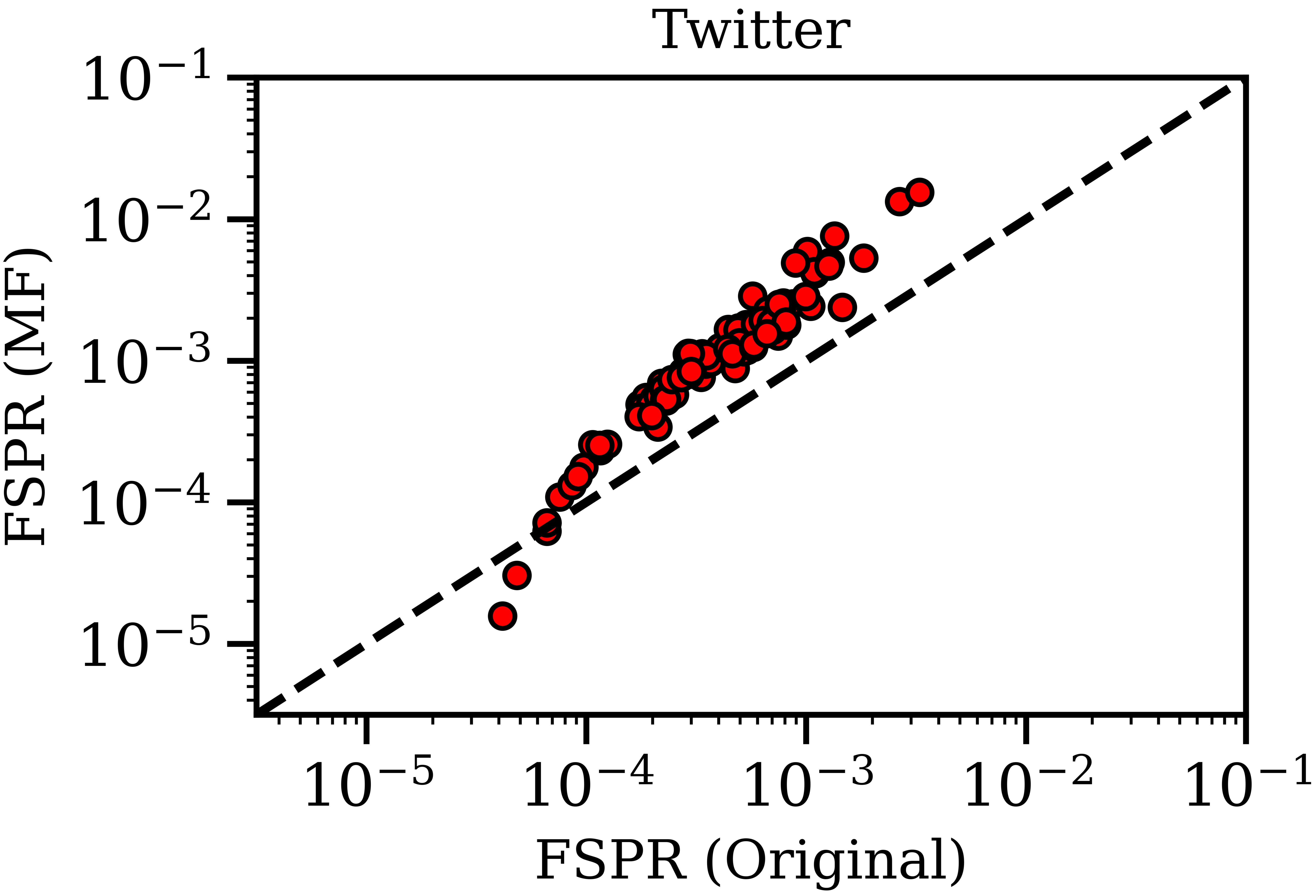}
\end{subfigure}
\vspace{2mm}
\begin{subfigure}{0.23\textwidth}
    \centering
    \includegraphics[width=\linewidth]{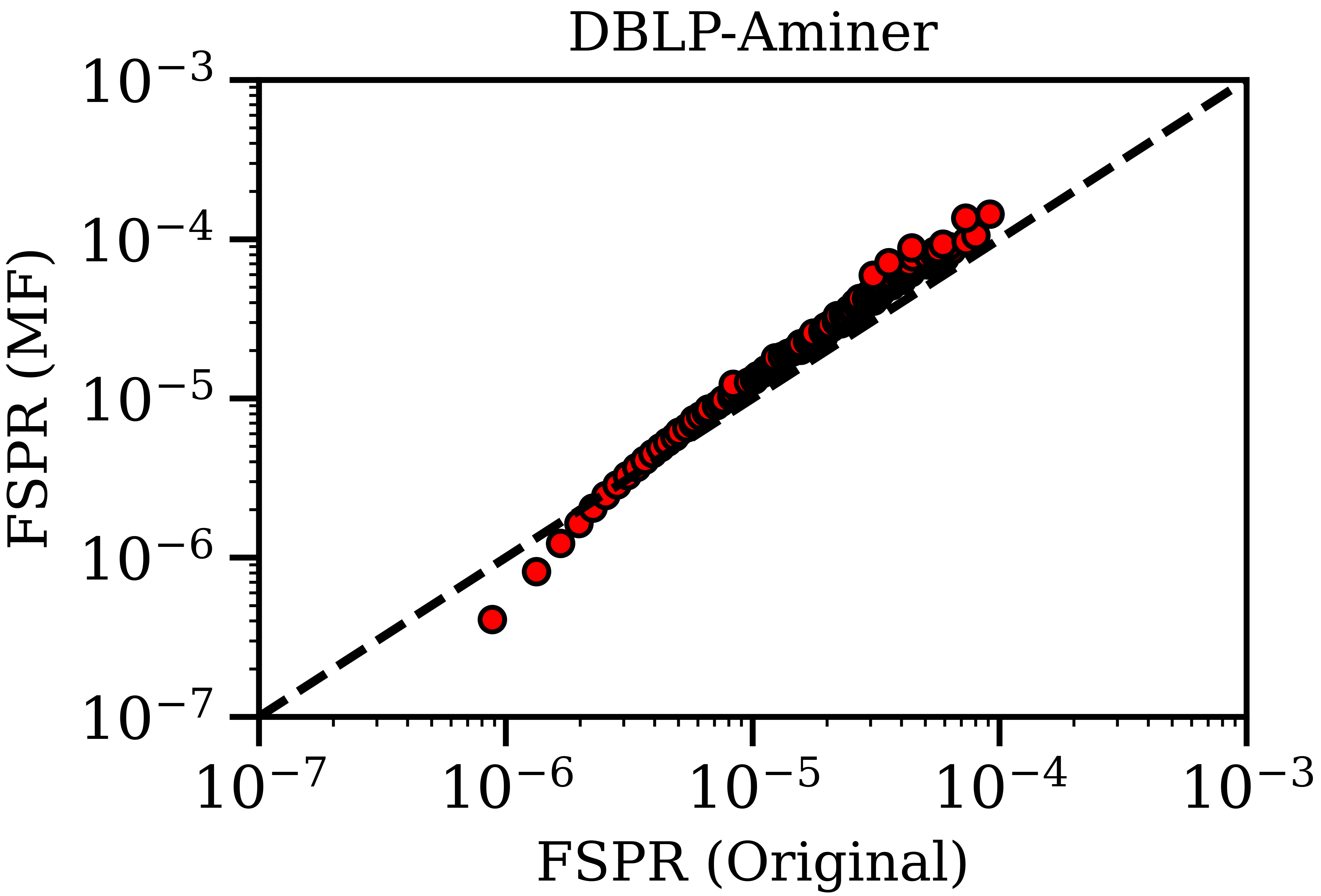}
\end{subfigure}
\begin{subfigure}{0.23\textwidth}
    \centering
    \includegraphics[width=\linewidth]{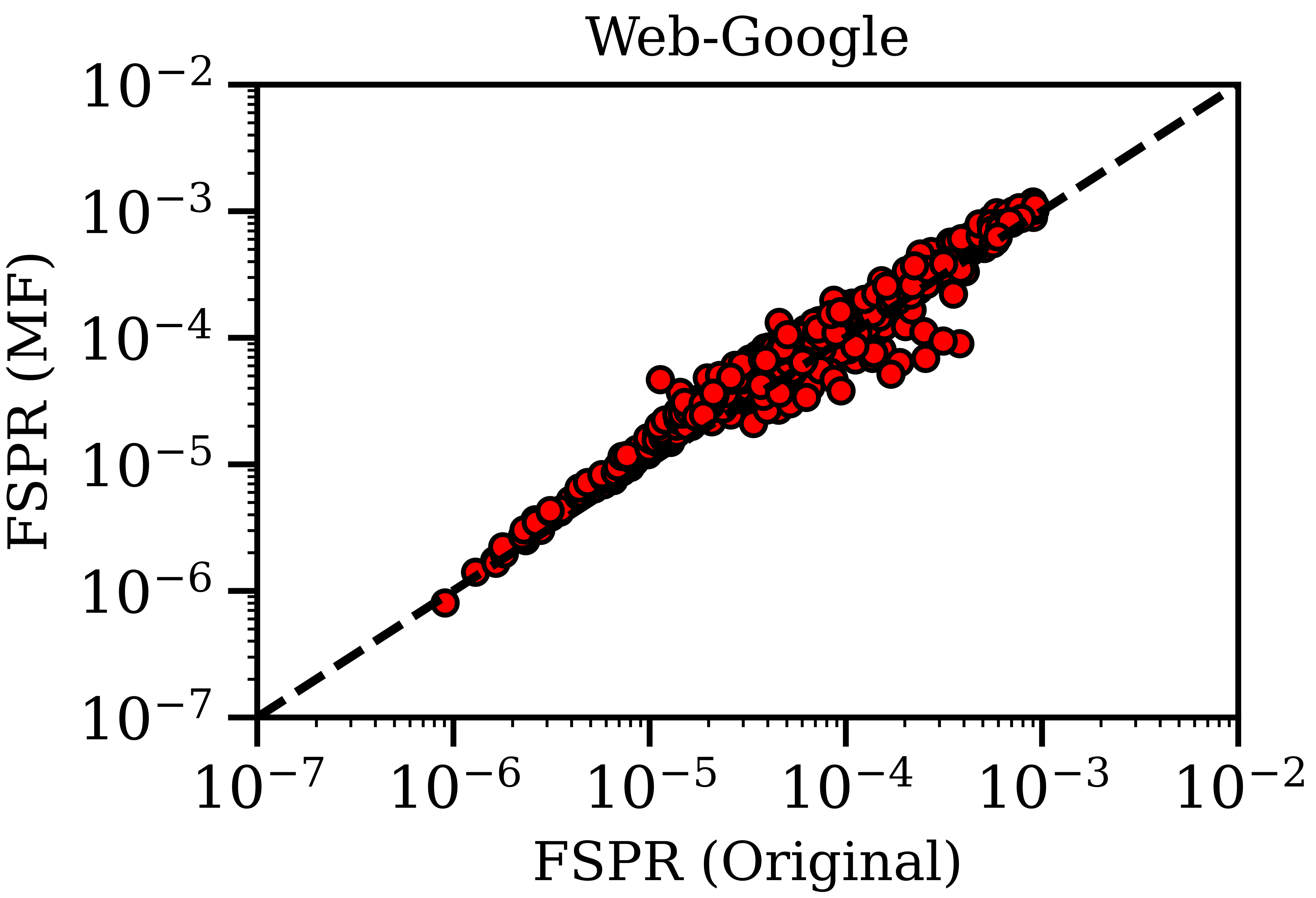}
\end{subfigure}
\begin{subfigure}{0.23\textwidth}
    \centering
    \includegraphics[width=\linewidth]{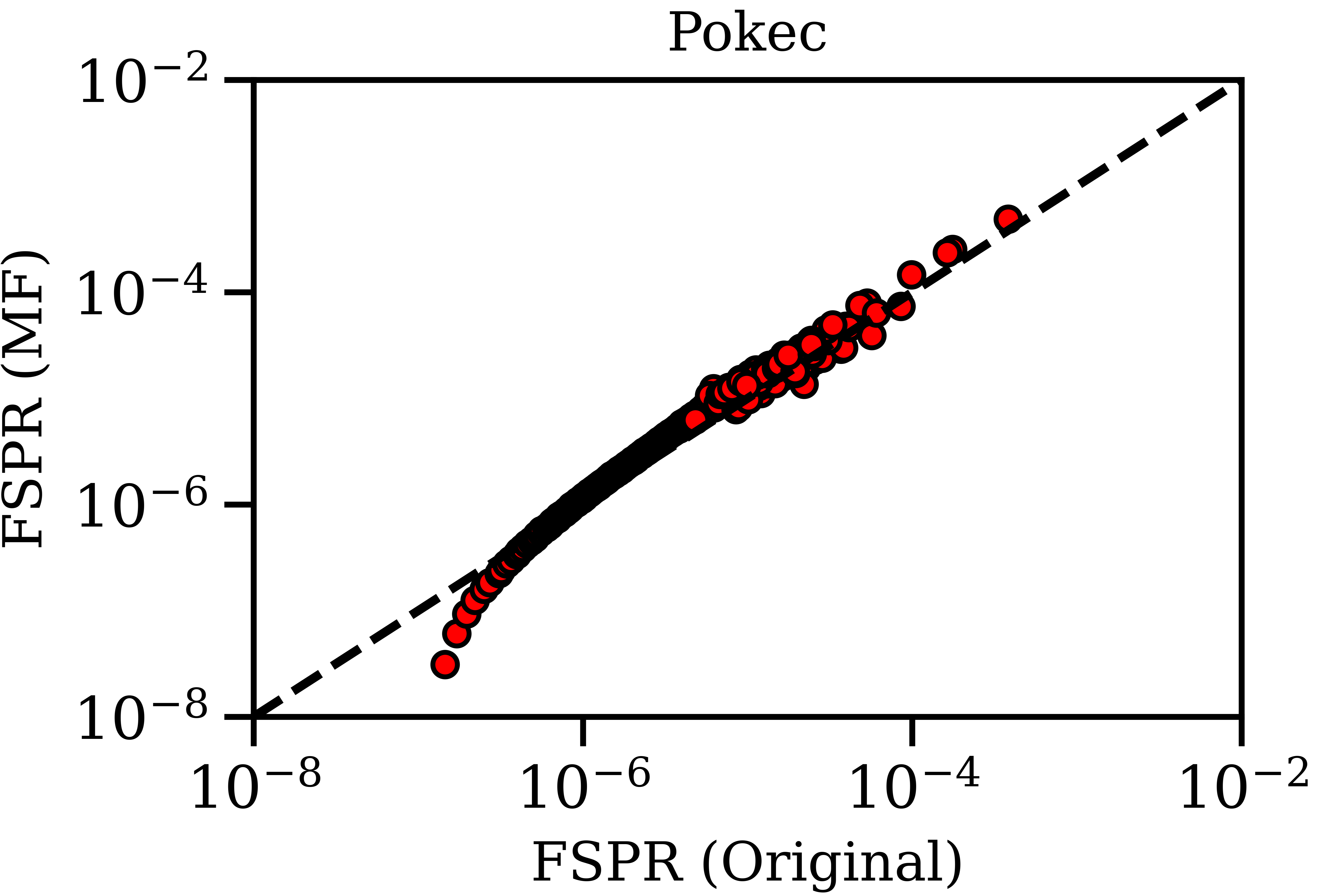}
\end{subfigure}
\begin{subfigure}{0.23\textwidth}
    \centering
    \includegraphics[width=\linewidth]{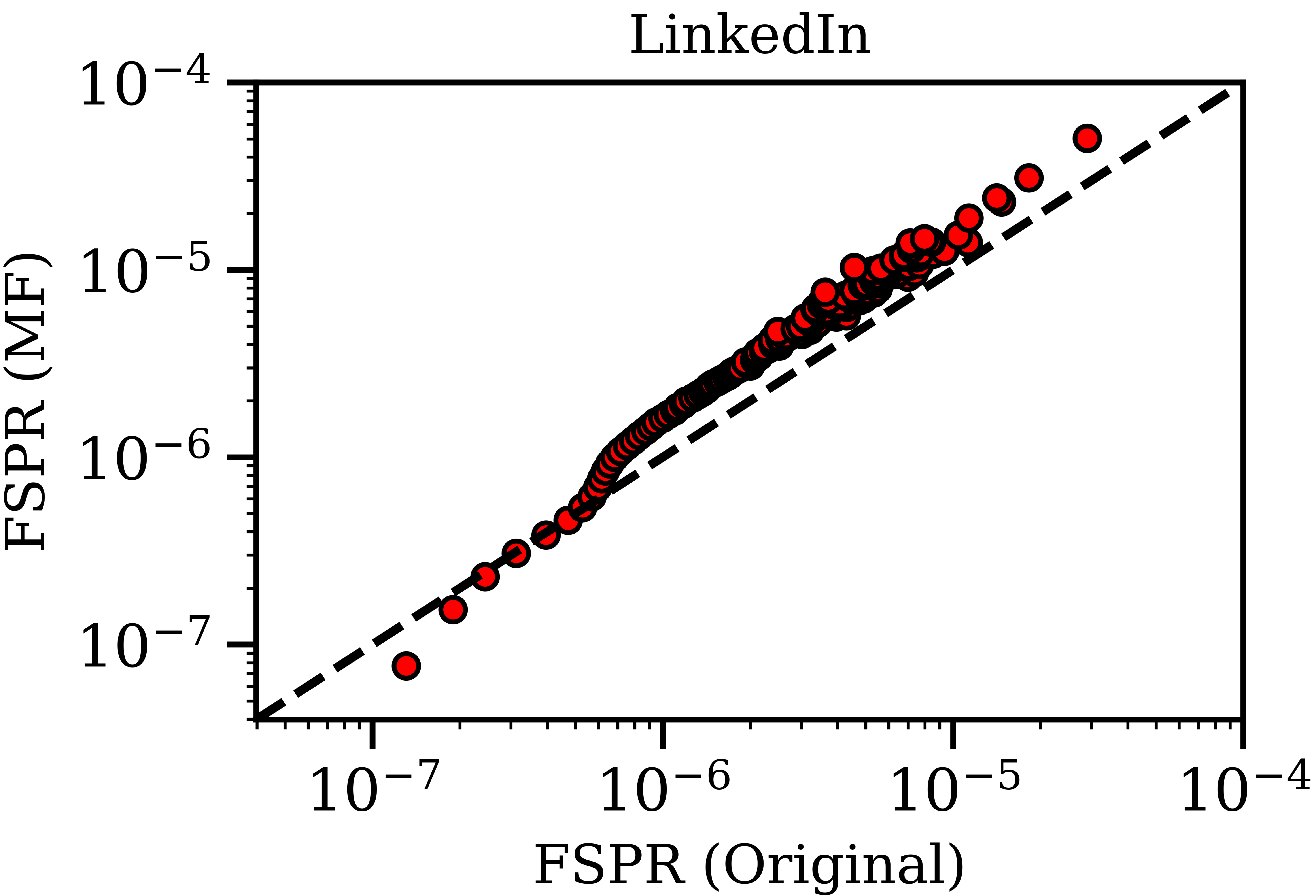}
\end{subfigure}
\caption{Mean field (MF) approximated FSPR is plotted against empirical average FSPR per degree-class.}
\label{est_vs_actual_fspr}
\end{figure}

Figure \ref{est_vs_actual_fspr} shows the mean-field approximated FSPR, i.e., computed using Eq.~(\ref{eq:equationB13}), versus original FSPR \cite{tsioutsiouliklis2021fairness}. Eq.~(\ref{eq:equationB13}) was solved using an iterative procedure of the FSPR algorithm. 
For each group compute the vector $p(k^{C}_{i})$, that was initialized uniformly $(p_{0}(k^{C}_{i}) = 1/N)$, and substituted into the right-hand side (RHS) of Eq.~(\ref{eq:equationB13}) and updated until convergence. The conditional probability $(P_{in}(k^{'} \mid k^{C}_{i}))$ is not an important component, as the summation in Eq.~(\ref{eq:equationB13}) implicitly captures the mean contribution of all predecessors of vertices degree, $k^{C}_{i}$. The algorithm converges rapidly, and empirical results in Figure \ref{est_vs_actual_fspr} show that the proposed mean-field approximation method provides excellent results. For large-scale datasets, including DBLP-Aminer, Web-Google, Pokec, and LinkedIn, explicit computation of the original FSPR is infeasible due to the memory constraints. Therefore, we use the GMRES-based iterative solver to obtain scalable and numerically consistent baseline FSPR scores (refer to Appendix~A for details). We also evaluated the correlation between the approximated and actual FSPR scores ($\tau$) using the Pearson correlation coefficient, as reported in Table \ref{tab:merged_results_reversed}. The results show correlations above 0.94 across all datasets, demonstrating the strong agreement and effectiveness of the proposed approximation.

\begin{figure}[t]
    \centering
\begin{subfigure}{0.23\textwidth}
    \centering
    \includegraphics[width=\linewidth]{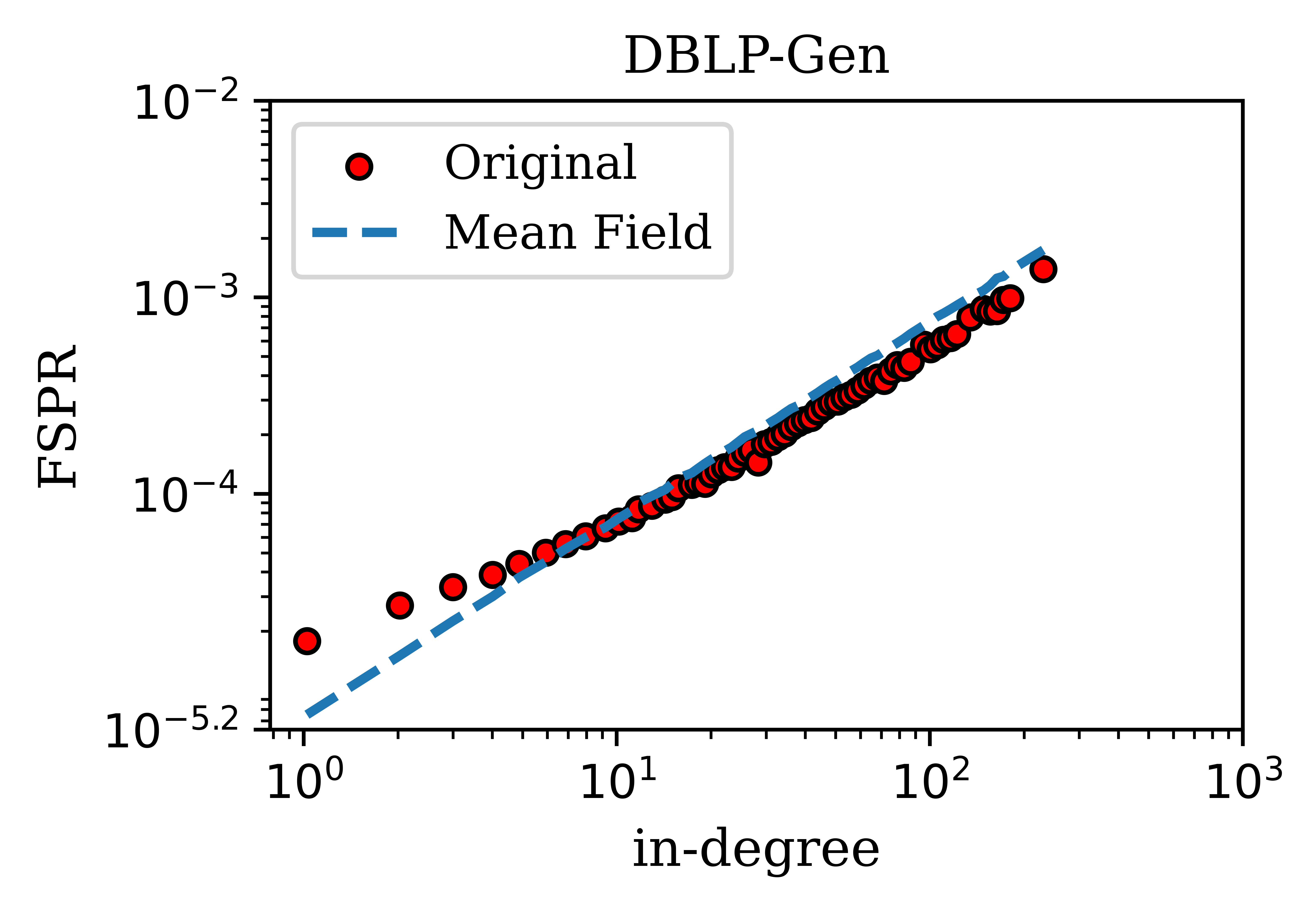}
\end{subfigure}
\vspace{2mm}
\begin{subfigure}{0.23\textwidth}
    \centering
    \includegraphics[width=\linewidth]{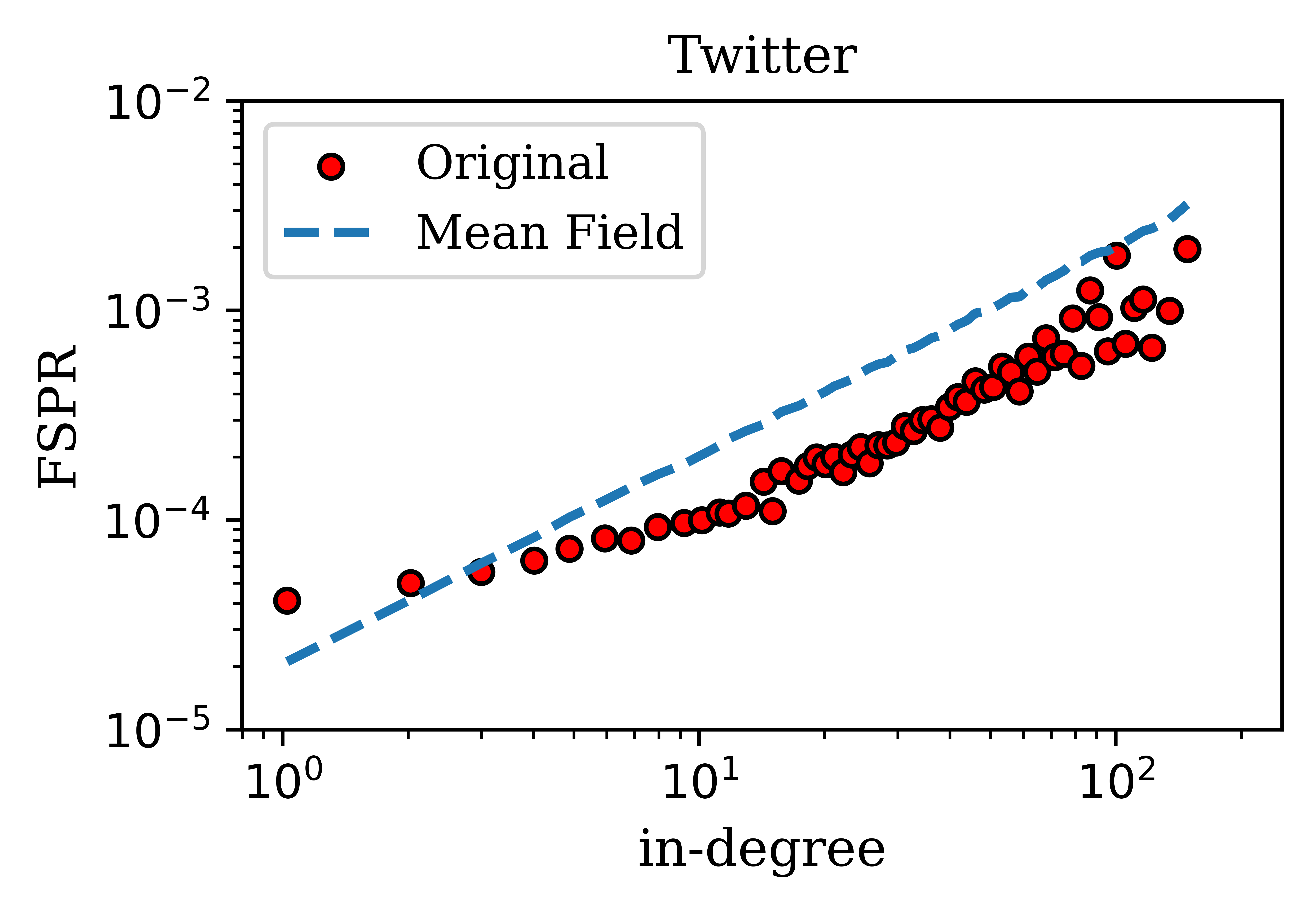}
\end{subfigure}
\vspace{2mm}
\begin{subfigure}{0.23\textwidth}
    \centering
    \includegraphics[width=\linewidth]{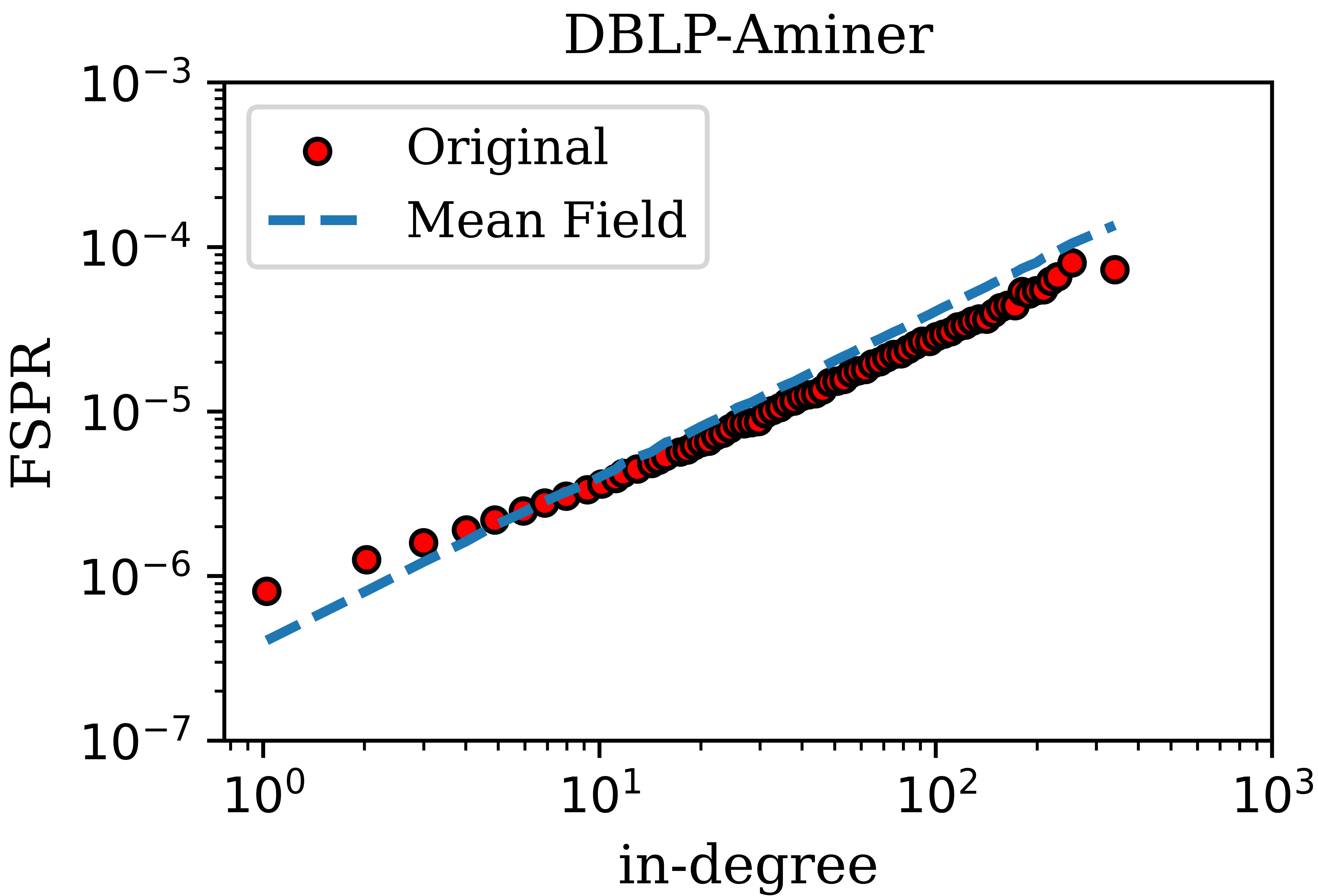}
\end{subfigure}
\begin{subfigure}{0.23\textwidth}
    \centering
    \includegraphics[width=\linewidth]{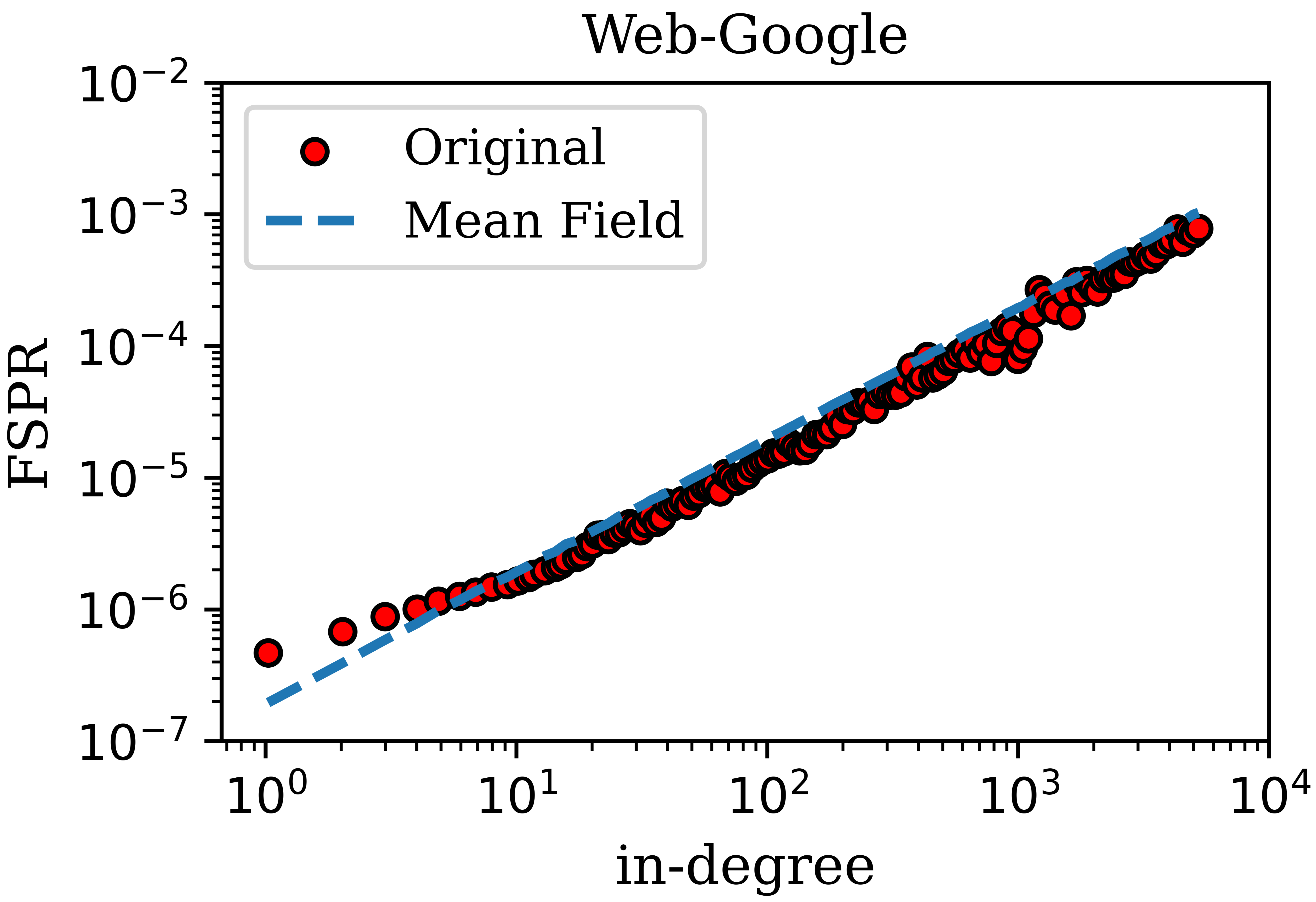}
\end{subfigure}
\begin{subfigure}{0.23\textwidth}
    \centering
    \includegraphics[width=\linewidth]{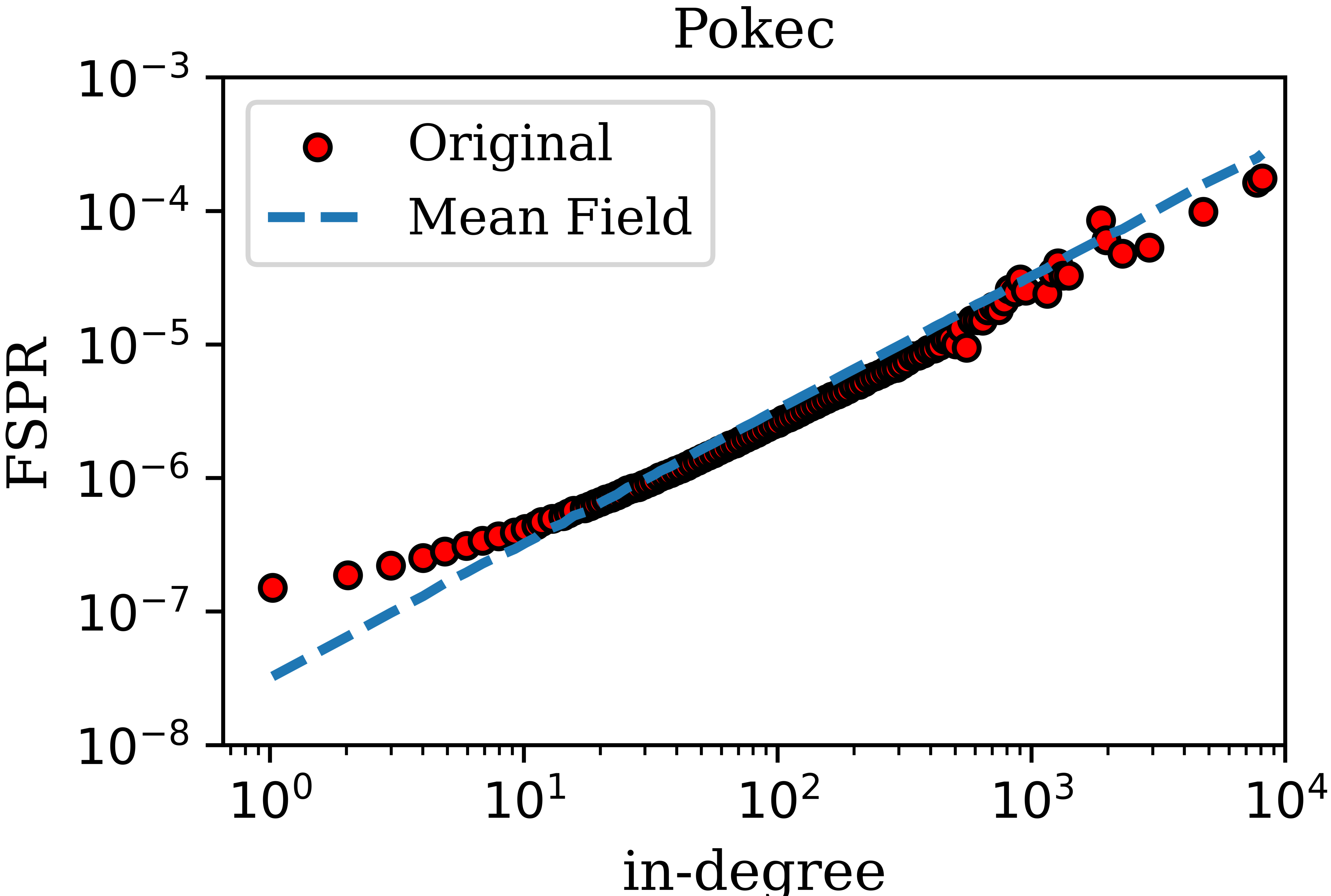}
\end{subfigure}
\begin{subfigure}{0.23\textwidth}
    \centering
    \includegraphics[width=\linewidth]{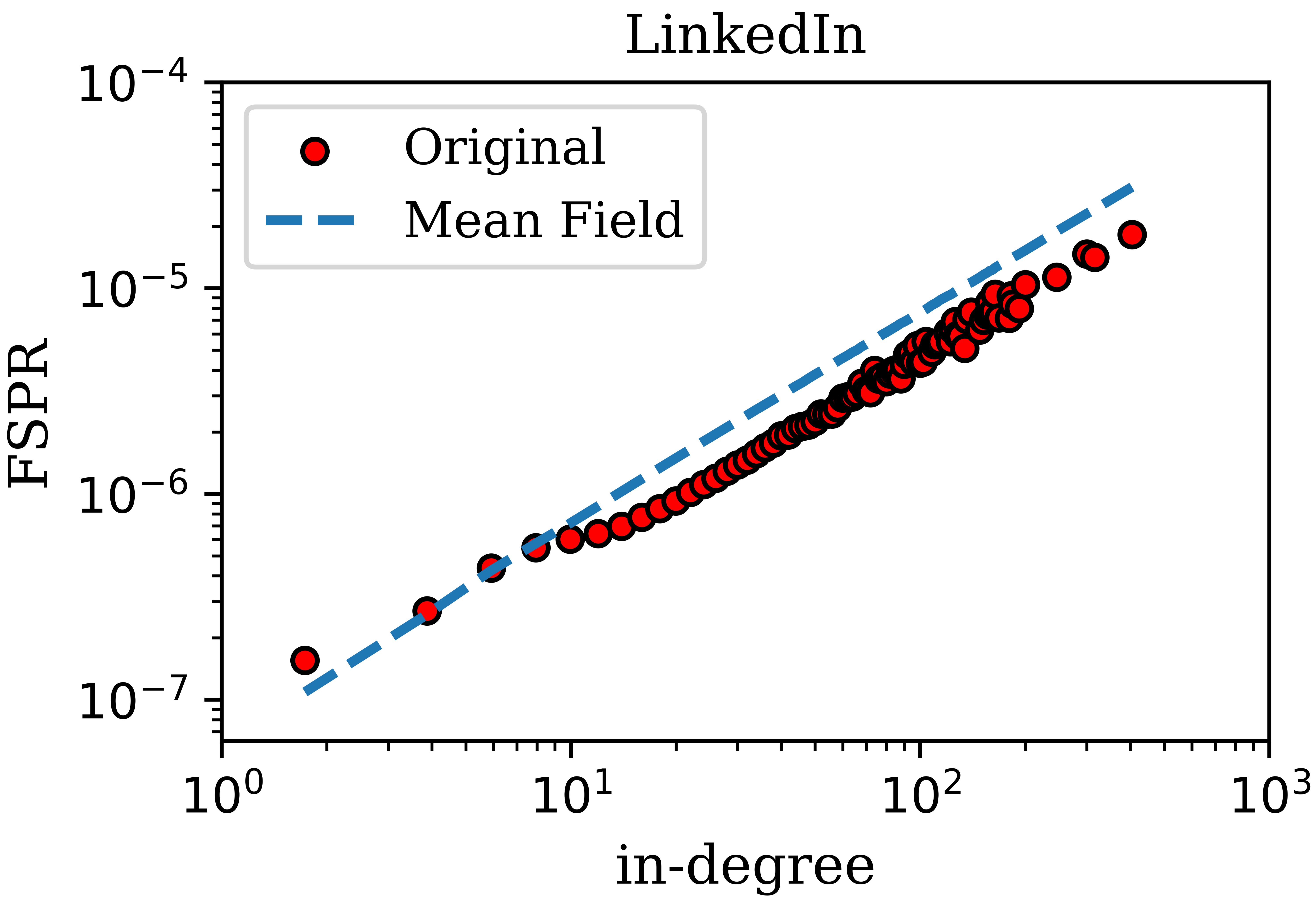}
\end{subfigure}
\caption{Fairness-Sensitive PageRank plotted against in-degree. The dashed blue line shows the approximation derived from Eq.~(\ref{eq:equationB15}).}
    \label{Fig:imageB}
\end{figure}

\begin{table}[t]
\centering
\caption{For each dataset, the table reports: (i) ${\rho}$: FSPR vs. in-degree correlation, (ii) $\tau$: MF approximated FSPR vs. original FSPR, (iii) UtilityLoss, and (iv) FairnessGap. Lower UtilityLoss and FairnessGap indicate better utility and fairness. Correlations are computed using the Pearson coefficient.}
\label{tab:merged_results_reversed}
\begin{tabular}{lccccc}
\toprule
\textbf{Dataset} & $\boldsymbol{\rho}$ & $\boldsymbol{\tau}$ & \textbf{UtilityLoss} $\downarrow$ & \textbf{FairnessGap} $\downarrow$ \\
\midrule
DBLP-Gen      & 0.94 & 0.99 & 0.000047  & 0.0048 \\
Twitter       & 0.85 & 0.94 & 0.000061  & 0.0192 \\
DBLP-Aminer   & 0.95 & 0.98 & 0.000002  & 0.0014 \\
Web-Google    & 0.90 & 0.97 & 0.000002  & 0.0008 \\
Pokec         & 0.94 & 0.99 & 0.000001  & 0.0001 \\
LinkedIn      & 0.89 & 0.98 & 0.000000  & 0.0003 \\
\bottomrule
\end{tabular}
\end{table}

We further examine the relationship between the FSPR and the in-degree of nodes. As shown in Table~\ref{tab:merged_results_reversed}, the correlation coefficient $(\rho)$ between FSPR and in-degree is consistently high across all datasets.
For small in-degree values, FSPR exhibits wide fluctuations within each degree class, and therefore, a direct plot of $\overline{p}(k^{C}_{\text{in}})$ versus in-degree does not reveal a clear trend. Therefore, we compute the average FSPR values across all nodes within logarithmic in-degree bins (multiplication factor 1.05), as shown in Figure~\ref{Fig:imageB}.
The empirical trend shows an approximately linear relationship between average FSPR and in-degree. Moreover, across all datasets, the mean FSPR closely follows the theoretical prediction given by Eq.~(\ref{eq:equationB15}) for both protected and unprotected groups.

\begin{figure}[t]
    \centering
\begin{subfigure}{0.23\textwidth}
    \centering
    \includegraphics[width=\linewidth]{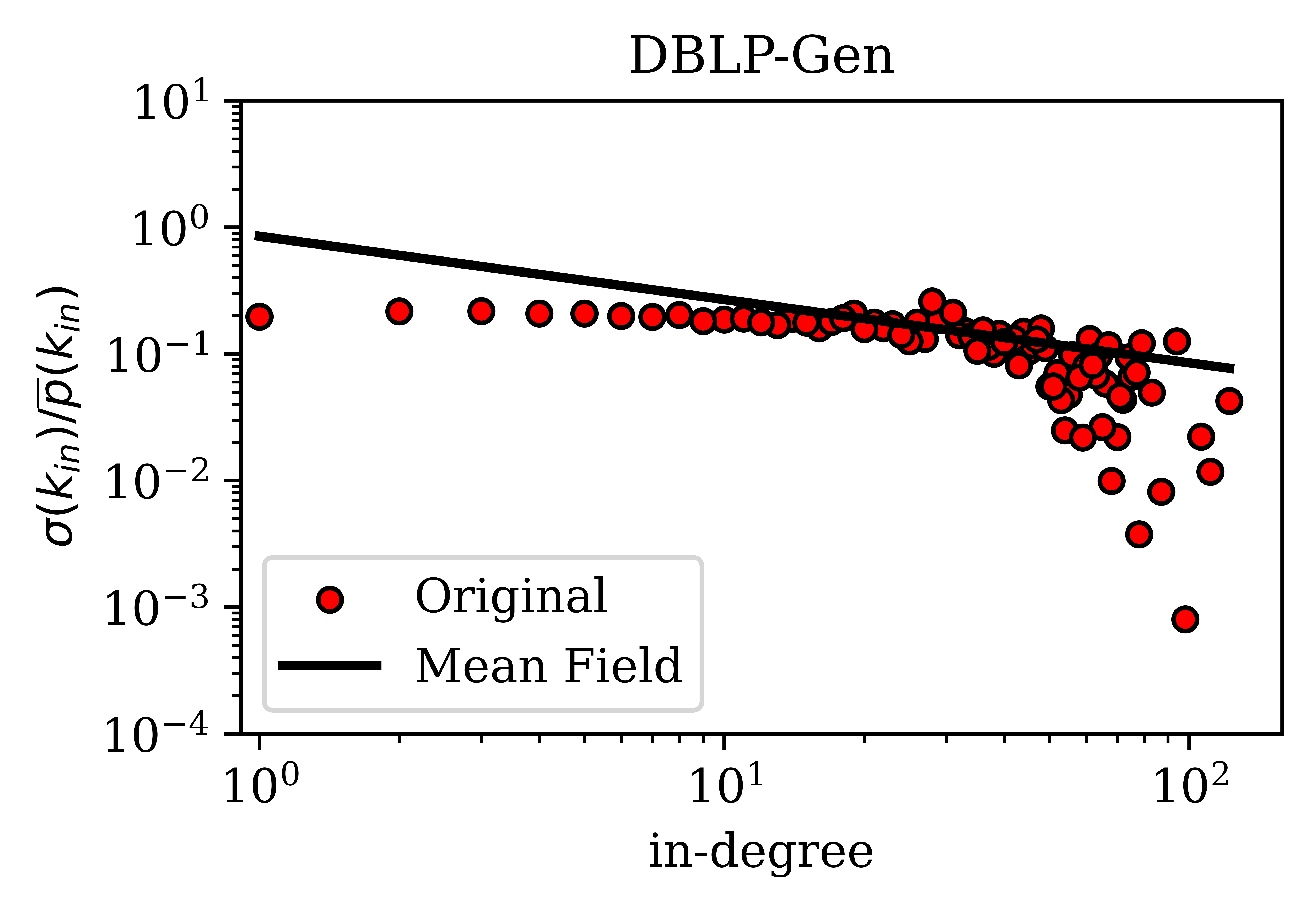}
\end{subfigure}
\vspace{2mm}
\begin{subfigure}{0.23\textwidth}
    \centering
    \includegraphics[width=\linewidth]{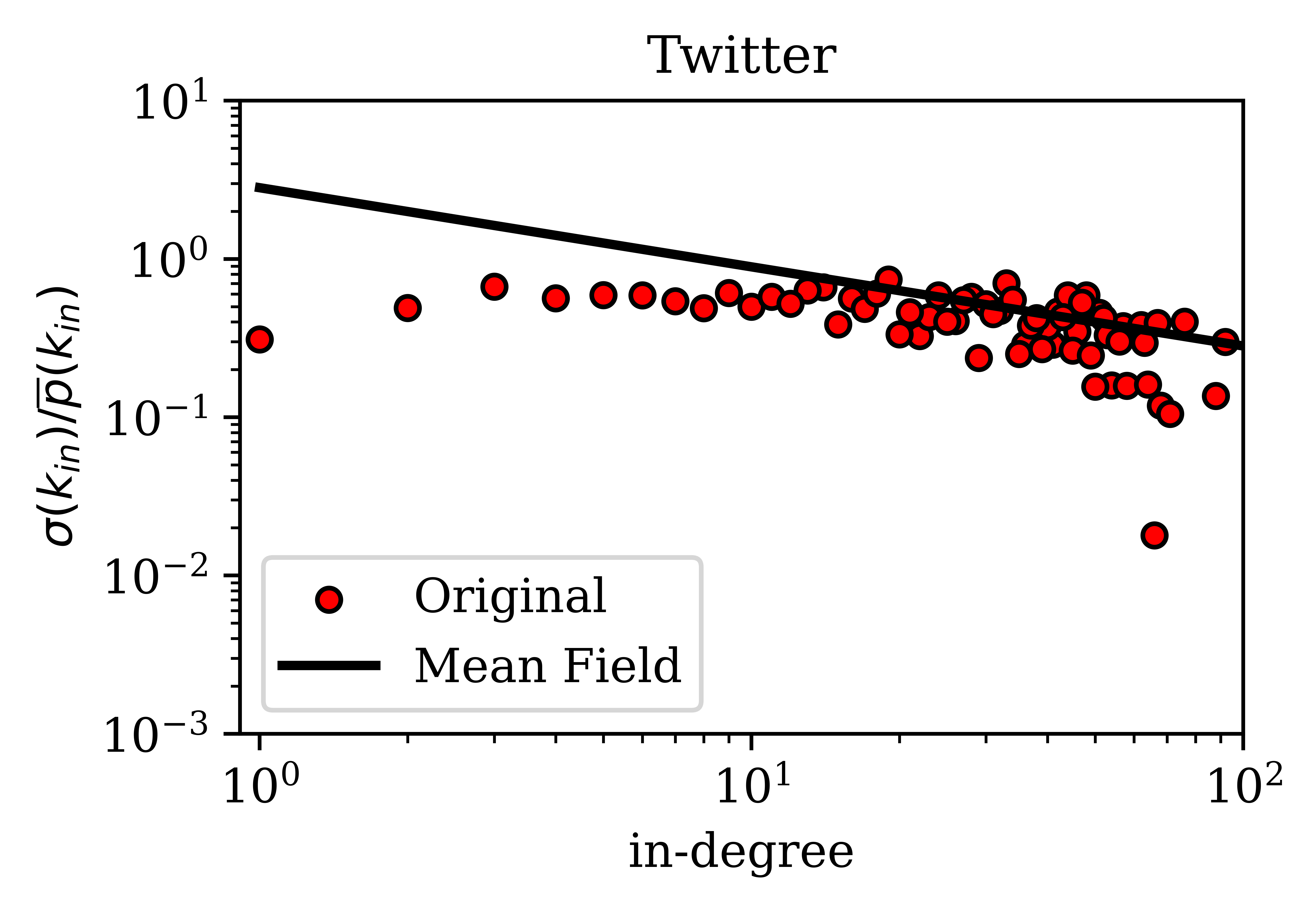}
\end{subfigure}
\vspace{2mm}
\begin{subfigure}{0.23\textwidth}
    \centering
    \includegraphics[width=\linewidth]{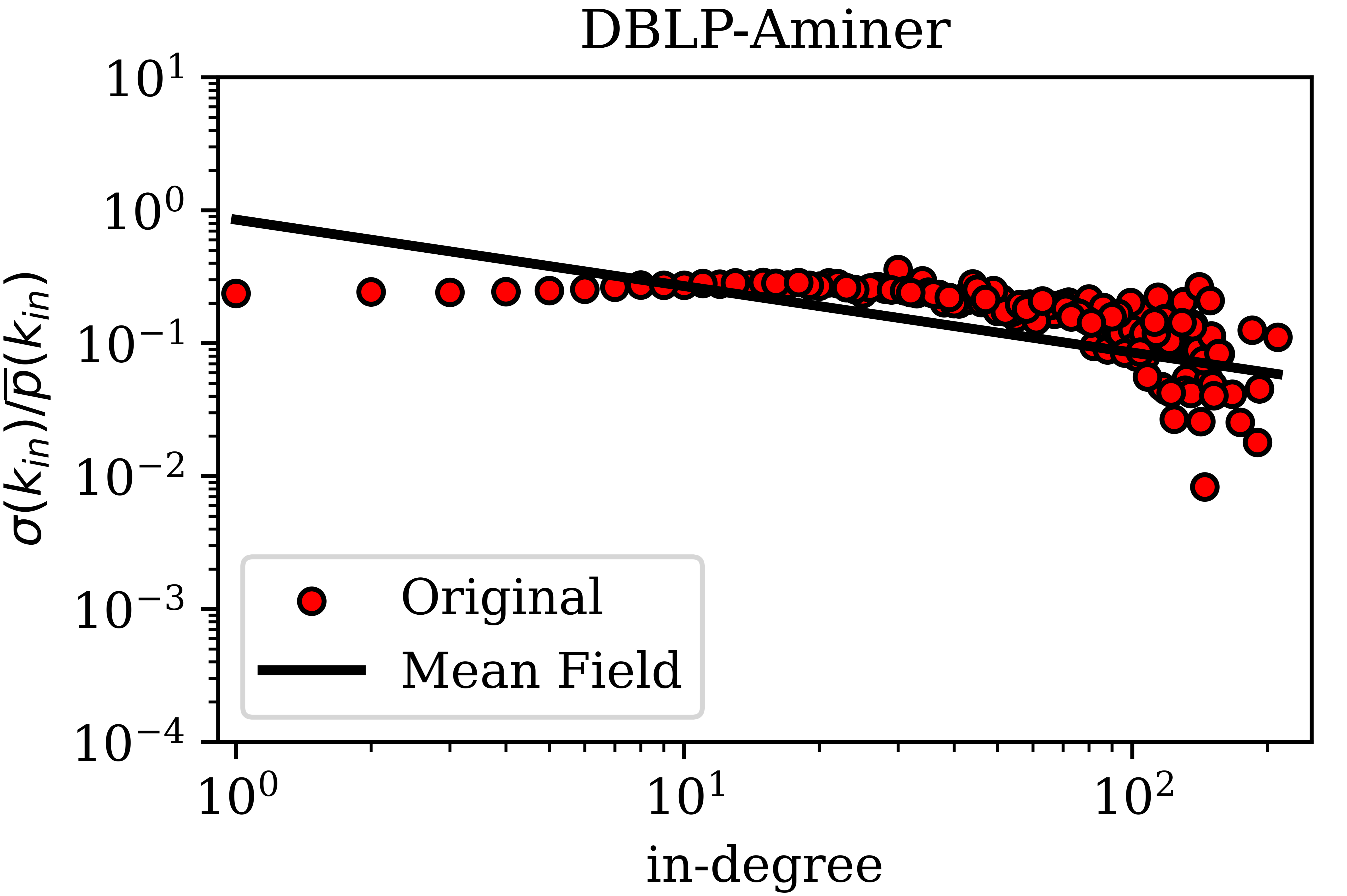}
\end{subfigure}
\begin{subfigure}{0.23\textwidth}
    \centering
    \includegraphics[width=\linewidth]{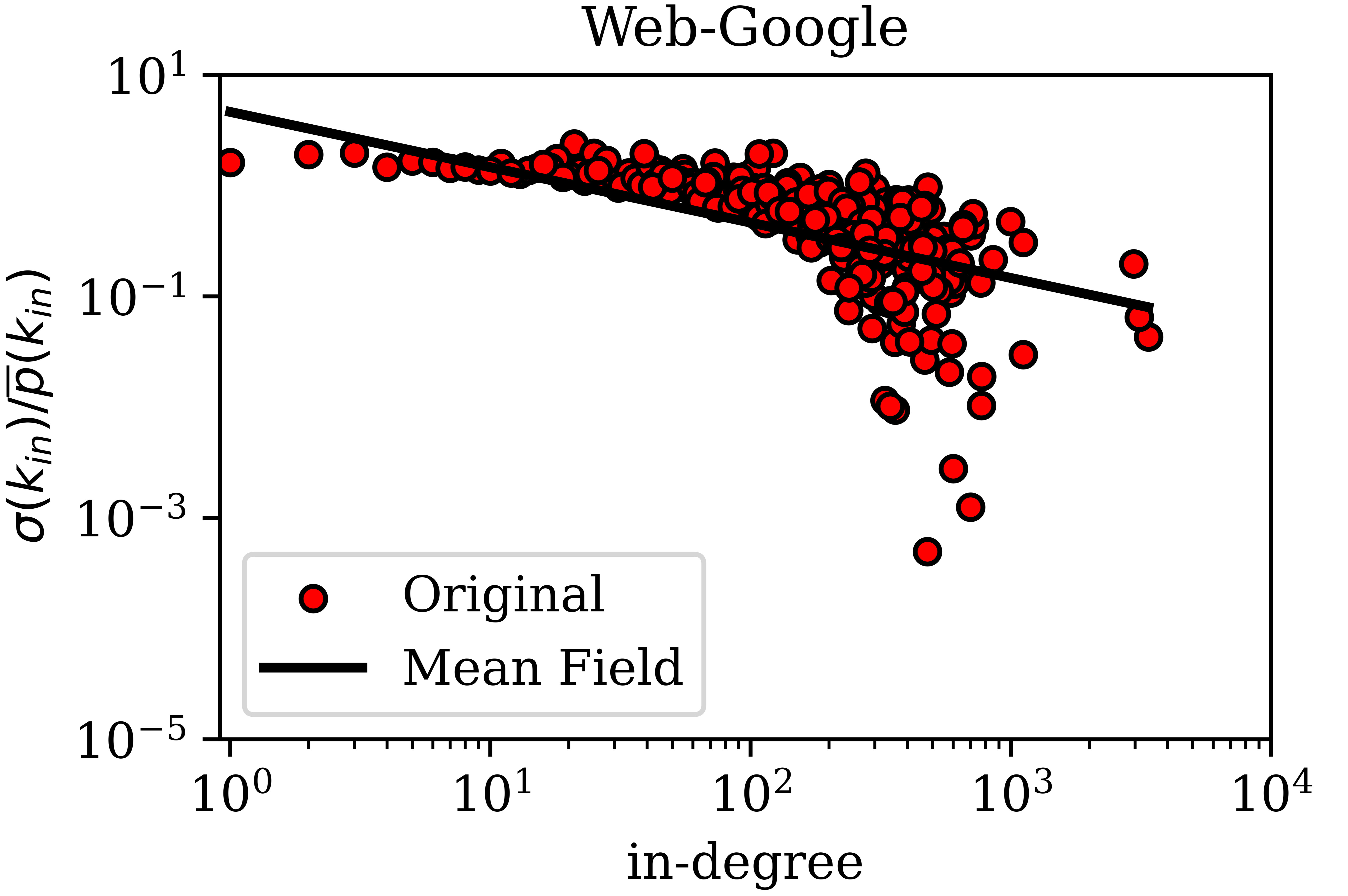}
\end{subfigure}
\begin{subfigure}{0.23\textwidth}
    \centering
    \includegraphics[width=\linewidth]{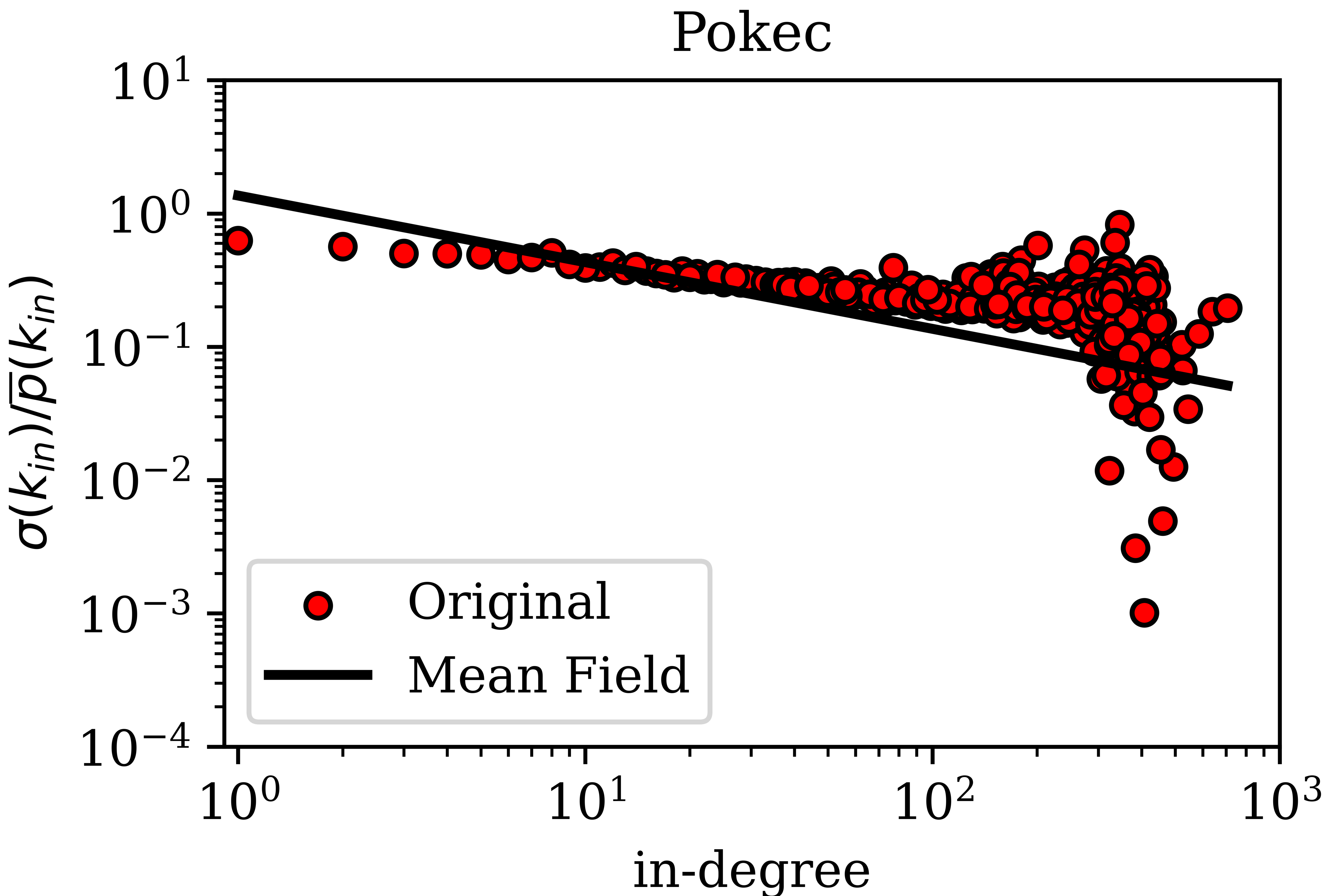}
\end{subfigure}
\begin{subfigure}{0.23\textwidth}
    \centering
    \includegraphics[width=\linewidth]{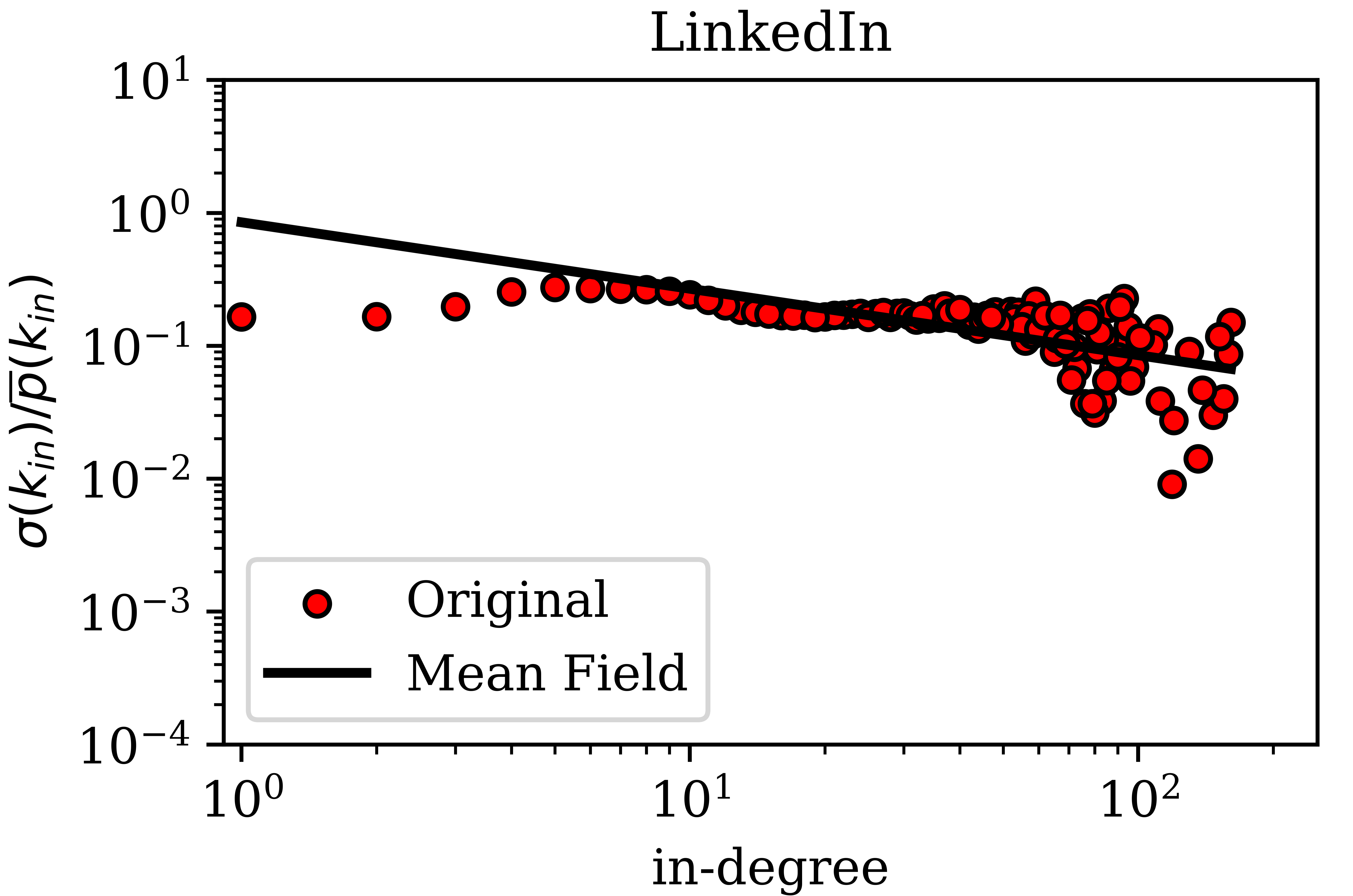}
\end{subfigure}
\caption{Coefficient of variation of FSPR with respect to in-degree.}
\label{Fig:imageC}
\end{figure}

Next, we empirically analyze the fluctuations of FSPR to analyze deviations from mean values. As discussed in Section~\ref{subsec:Fluctuation}, nodes with low in-degree exhibit higher variability due to large second moments of the degree distribution, whereas fluctuations diminish as in-degree $k^{C}_{i,\text{in}}$ increases. Figure~\ref{Fig:imageC} shows the coefficient of variation \( \sigma(k^{C}_{i,\text{in}}) / \overline{p}(k^{C}_{i,\text{in}}) \) with respect to the \(k^{C}_{i,\text{in}}\) (in-degree), averaged over out-degree. Empirical results demonstrate a decreasing trend as $k^{C}_{i,\text{in}}$  grows. Overall, the mean-field approximation closely aligns with the empirical observations.

We finally assess the proposed approximation using two complementary metrics: UtilityLoss and FairnessGap. UtilityLoss measures the fidelity of the approximation by how closely the approximate FSPR scores $\hat{p}$ match the exact FSPR scores $p$, and is defined as the mean absolute deviation between them.
\begin{equation*}
    \text{UtilityLoss} = \frac{1}{N}\sum_{i=1}^{N}|\hat{p}(i)-p(i)|
\end{equation*}
Low UtilityLoss values indicate that the approximate FSPR scores closely match the original FSPR scores.

The FairnessGap metric quantifies the deviation from the desired PageRank mass for each group and is defined as:
\begin{equation*}
\text{FairnessGap}(C) = \left|\sum_{i\in T}\hat{p}(i) - \phi_{C} \right|
\end{equation*}
where $C \in \{P,U\}$ denotes the protected or unprotected group, and $\phi_{C}$ is the target PageRank mass for group $C$. In the case of two groups, FairnessGap will be the same for both groups. Low FairnessGap values indicate that the approximate FSPR scores more closely satisfy the fairness constraint.

Table \ref{tab:merged_results_reversed} shows that, across all datasets, the proposed approximation method exhibits consistently low UtilityLoss (typically $\leq 10^{-4}$) and very small FairnessGap, preserving both the node-level PageRank scores and the group-level fairness.

\subsection{Scalability Analysis} 
To assess the practical scalability of the proposed mean-field approximation, we compare its runtime performance with the original Fairness-Sensitive PageRank (FSPR) method and the GMRES-based iterative solver used for large-scale networks, since the original FSPR method returned an out-of-memory error. All methods are implemented in the same computational environment and evaluated using identical hardware configurations to ensure a fair comparison. The experiments are performed on a workstation running Microsoft Windows $11$ Pro, equipped with a $13\text{-th}$ Generation Intel Core $i7-13700$ CPU ($2.1$GHz, $16$ cores, $24$ threads) and $32$GB of RAM.

Table~\ref{tab:runtime-comparison} summarizes the runtime of the exact FSPR formulation, the GMRES-based iterative solver, and the proposed mean-field approximation across datasets of increasing scale. The exact FSPR formulation requires dense matrix construction and manipulation, resulting in cubic-time complexity and quadratic memory usage (Section~\ref{sec:fspr_MFA_complexity}), which makes it infeasible for moderately large networks. This is confirmed empirically by out-of-memory failures on datasets such as DBLP-Aminer, Pokec, LinkedIn, and Web-Google.

To handle large networks, we therefore rely on the GMRES-based solver, which avoids explicit matrix inversion and operates on sparse graph structures. While this significantly improves scalability, its runtime depends on both the number of outer fairness optimization steps and inner Krylov subspace iterations (Appendix~\ref{sec:GMRES_complexity}), leading to rapidly increasing runtimes as network size grows, particularly with the number of edges.

In contrast, the proposed mean-field approximation exhibits consistently low runtime across all datasets. Since the method requires only a single pass over the edge set to compute node in-degrees and group-level aggregates, its runtime scales linearly with the number of edges, consistent with the theoretical analysis in Section~\ref{sec:Mean_field_Complexity}.

\begin{table}[t]
\caption{Runtime (in seconds) for original FSPR, GMRES-based solver, and the proposed Mean-Field Approximation method across datasets of varying scale. ``OOM'' denotes out-of-memory failure.}
\centering
\small
\begin{tabular}{lrrr}
\toprule
\textbf{Dataset} & \textbf{FSPR} & \textbf{GMRES} & \textbf{Mean-Field} \\
\midrule
Dblp-Gen       & 1328.91  & 0.62     & 0.0206 \\
Twitter        & 3986.99  & 25.07    & 0.0248 \\
Dblp-Aminer    & OOM      & 169.40   & 0.695  \\
Web-Google     & OOM      & 151.55   & 1.15   \\
Pokec          & OOM      & 4066.07  & 3.40   \\
LinkedIn       & OOM      & 2343.21  & 6.76   \\
\bottomrule
\end{tabular}
\label{tab:runtime-comparison}
\end{table}

\section{Conclusion}
This work presents an efficient mean-field approximation method for Fairness-Sensitive PageRank that addresses the scalability limitations of existing fairness-aware ranking methods. By introducing a degree-class representation and deriving a fair mean-field formulation, the proposed method eliminates the need for solving high-dimensional convex optimization problems while preserving group-level fairness guarantees. We further analyzed intra-class fluctuations through a variance formulation, providing insight into the reliability of the approximation across different degree regimes. Extensive experiments on real-world networks demonstrated that the mean-field approximation closely matches exact FSPR scores with substantially lower computational cost for large-scale networks.

\appendices

\section{GMRES-based Solver for Baseline Computation}
To evaluate the mean-field approximation, we require baseline Fairness-Sensitive PageRank (FSPR) scores. However, the original FSPR formulation requires constructing the matrix  
\[
Q = \gamma [I - (1-\gamma)P]^{-1}
\]
which leads to $\mathcal{O}(N^2)$ memory and $\mathcal{O}(N^3)$ time complexity. Consequently, the FSPR computation is feasible only for small-scale networks and becomes computationally infeasible for medium- and large-scale networks. 
 
Therefore, for moderate and large networks, we adopt a scalable solver based on the Generalized Minimal Residual (GMRES) method \cite{saad1986gmres}, which implicitly applies the action of $Q$ by solving sparse linear systems, without explicitly forming the matrix. The teleportation (jump) vector is updated iteratively to enforce the fairness constraint, enabling the solver to maintain high accuracy while scaling to large networks.

\begin{table*}[t]
\caption{Comparison between the original FSPR solver and the scalable GMRES-based solver. Lower values of protected mass $(\Delta)$, $L_1$, and $L_2$ distances indicate closer numerical agreement with the original FSPR solution, whereas higher values of Kendall’s $\tau$ and top-$k$ overlap indicate stronger ranking consistency.}
\centering
\small
\begin{tabular}{lccccccc}
\toprule
\textbf{Dataset} & \textbf{\#Nodes} & \textbf{\#Edges} & $\Delta$ Protected Mass $\downarrow$ & $L_1 \downarrow$ & $L_2 \downarrow$ & Kendall $\tau \uparrow$ & Top-200 $\uparrow$ \\
\midrule
Twitch-DE \cite{rozemberczki2019multiscale}    & 	9498 & 153138 & $1.08{\times}10^{-3}$ & $3.15{\times}10^{-2}$ & $1.09{\times}10^{-3}$ & 0.933 & 0.985 \\
Twitch-ENGB \cite{rozemberczki2019multiscale}   & 7126 & 35324 & $2.78{\times}10^{-4}$ & $1.69{\times}10^{-2}$ & $4.15{\times}10^{-4}$ & 0.968 & 0.990 \\
Dblp-Course \cite{tsioutsiouliklis2020fairness}  & 13015 & 79972 & $7.18{\times}10^{-5}$ & $1.05{\times}10^{-2}$ & $1.42{\times}10^{-4}$ & 0.984 & 0.990 \\
Books \cite{tsioutsiouliklis2021fairness}         & 92 & 748 & $3.97{\times}10^{-3}$ & $7.95{\times}10^{-3}$ & $8.30{\times}10^{-4}$ & 0.987 & 0.980 \\
Twitter \cite{tsioutsiouliklis2021fairness}      & 18470 & 61157 & $3.17{\times}10^{-3}$ & $2.15{\times}10^{-2}$ & $3.54{\times}10^{-4}$ & 0.952 & 0.970 \\ 
Dblp-Gen \cite{tsioutsiouliklis2022link}   & 16501 & 66613 &
$2.46{\times}10^{-5}$ &
$7.1{\times}10^{-3}$  &
$8.66{\times}10^{-5}$ & 0.987 &
0.995 \\
Dblp-Pub \cite{tsioutsiouliklis2022link}   & 16501 & 66613 &
$3.75{\times}10^{-9}$ &
$2.2{\times}10^{-2}$  &
$2.6{\times}10^{-4}$  & 0.967 &
0.98 \\
\bottomrule
\end{tabular}
\label{tab:fairpr-validation}
\end{table*}

\subsection*{Complexity of the GMRES-Based Solver}\label{sec:GMRES_complexity}
The computational complexity of the GMRES-based approach depends on both the outer fairness optimization procedure and the inner Krylov subspace iterations.

\subsubsection{Time Complexity} 
Each GMRES iteration requires one sparse matrix–vector multiplication involving the PageRank transition matrix $P$, which incurs $\mathcal{O}(M)$ time, where $M(=|E|)$ denotes the number of edges in the network. Let $T_{opt}$ be the number of outer optimization iterations and $T_{GMRES}$ the number of GMRES iterations per optimization step. The overall time complexity is
\[\mathcal{O}(T_{opt}.T_{GMRES}.M)\]
Thus, while the scalability of the approach is primarily driven by the graph's sparsity, its overall runtime is governed by both the optimization dynamics and the convergence behavior of the Krylov subspace method.

\subsubsection{Space Complexity}
The memory requirements of the GMRES-based solver consist of three main components. 

The memory requirements of the GMRES-based solver consist of three primary components. First, storing the sparse graph requires $\mathcal{O}(N+M)$ memory, where $N$ and $M$ are the numbers of nodes and edges, respectively. Second, storing the PageRank vector and the teleportation vector incurs an additional $\mathcal{O}(N)$ memory cost.
Finally, GMRES maintains a set of Krylov subspace basis vectors during the iterative solution process. Since $T_{GMRES}$ basis vectors of dimension $N$ are stored, this contributes an additional $\mathcal{O}(T_{GMRES}.N)$ memory overhead. Therefore, the overall memory complexity is 
\[\mathcal{O}(N+M+T_{GMRES}.N)\]

The GMRES-based solver avoids explicit matrix inversion and operates directly on sparse graph structures, thereby substantially improving scalability compared to the original FSPR formulation. However, since GMRES is embedded within an outer fairness optimization loop, the overall time complexity scales as $\mathcal{O}(T_{opt}.T_{GMRES}.M)$, and the memory complexity is $\mathcal{O}(N+M+T_{GMRES}.N)$ due to storage of Krylov subspace vectors. 
In contrast, the proposed mean-field approximation eliminates both the outer optimization loop and the Krylov subspace computations. By relying solely on node in-degrees and group-level aggregate statistics, the method achieves a strictly linear time and space complexity. Consequently, the mean-field approximation offers superior scalability and predictable linear performance compared to the GMRES-based solver.

\subsection*{Empirical Validation of GMRES-Based Solver}
We compare the scalable GMRES-based solver with the original FSPR solver on datasets where the exact computation of FSPR is feasible. The evaluation considers the following complementary metrics.

\textbf{Protected mass deviation} measures the difference between the PageRank mass assigned to the protected group and the target value $\phi$. Lower values indicate that the GMRES-based solver closely achieves the target mass.

\textbf{$L_{1}$ and $L_{2}$ distances} measure numerical differences between the FSPR scores produced by the exact solver and GMRES-based solver; lower values indicate higher approximation fidelity.

\textbf{Kendall's rank correlation} measures the degree of agreement between the nodes' ranking orders induced by the GMRES-based and exact solvers.

\textbf{Top-\(K\) influential node overlap} measures the consistency among the top-\(K\) highest-ranked nodes.

The results in Table \ref{tab:fairpr-validation} demonstrate that the GMRES-based solver closely approximates the exact FSPR scores on all considered datasets for which exact computation is possible. The protected mass deviation remains consistently very small, indicating that the fairness constraint is accurately satisfied. The $L_{1}$ and $L_{2}$ distances are low across all datasets, indicating strong numerical agreement between the GMRES-based and exact FSPR solutions. Moreover, ranking consistency is preserved, with Kendall’s $\tau$ exceeding $0.93$ and $Top-200$ overlapping above 0.97 across all cases. Collectively, these results validate the correctness and reliability of the GMRES-based solver as a scalable alternative to the exact FSPR formulation for moderate- and large-scale networks.  


\balance
\bibliographystyle{IEEEtran}
\bibliography{sn-bibliography}

\begin{thebibliography}{10}
\providecommand{\url}[1]{#1}
\csname url@samestyle\endcsname
\providecommand{\newblock}{\relax}
\providecommand{\bibinfo}[2]{#2}
\providecommand{\BIBentrySTDinterwordspacing}{\spaceskip=0pt\relax}
\providecommand{\BIBentryALTinterwordstretchfactor}{4}
\providecommand{\BIBentryALTinterwordspacing}{\spaceskip=\fontdimen2\font plus
\BIBentryALTinterwordstretchfactor\fontdimen3\font minus \fontdimen4\font\relax}
\providecommand{\BIBforeignlanguage}[2]{{%
\expandafter\ifx\csname l@#1\endcsname\relax
\typeout{** WARNING: IEEEtran.bst: No hyphenation pattern has been}%
\typeout{** loaded for the language `#1'. Using the pattern for}%
\typeout{** the default language instead.}%
\else
\language=\csname l@#1\endcsname
\fi
#2}}
\providecommand{\BIBdecl}{\relax}
\BIBdecl

\bibitem{saxena2024fairsna}
A.~Saxena, G.~Fletcher, and M.~Pechenizkiy, ``Fairsna: Algorithmic fairness in social network analysis,'' \emph{ACM Computing Surveys}, vol.~56, no.~8, pp. 1--45, 2024.

\bibitem{dimaggio2012network}
P.~DiMaggio and F.~Garip, ``Network effects and social inequality,'' \emph{Annual review of sociology}, vol.~38, no.~1, pp. 93--118, 2012.

\bibitem{saxena2025homophily}
A.~Saxena, G.~Kumar, and C.~Meena, ``Homophily in complex networks: Measures, models, and applications,'' \emph{arXiv preprint arXiv:2509.18289}, 2025.

\bibitem{stoica2024fairness}
A.-A. Stoica, N.~Litvak, and A.~Chaintreau, ``Fairness rising from the ranks: Hits and pagerank on homophilic networks,'' in \emph{Proceedings of the ACM Web Conference 2024}, 2024, pp. 2594--2602.

\bibitem{saxena2022hm}
A.~Saxena, G.~Fletcher, and M.~Pechenizkiy, ``Hm-eiict: Fairness-aware link prediction in complex networks using community information,'' \emph{Journal of Combinatorial Optimization}, vol.~44, no.~4, pp. 2853--2870, 2022.

\bibitem{dong2023fairness}
Y.~Dong, J.~Ma, S.~Wang, C.~Chen, and J.~Li, ``Fairness in graph mining: A survey,'' \emph{IEEE Transactions on Knowledge and Data Engineering}, vol.~35, no.~10, pp. 10\,583--10\,602, 2023.

\bibitem{brin1998anatomy}
S.~Brin and L.~Page, ``The anatomy of a large-scale hypertextual web search engine,'' \emph{Computer networks and ISDN systems}, vol.~30, no. 1-7, pp. 107--117, 1998.

\bibitem{gleich2015pagerank}
D.~F. Gleich, ``Pagerank beyond the web,'' \emph{siam REVIEW}, vol.~57, no.~3, pp. 321--363, 2015.

\bibitem{tsioutsiouliklis2021fairness}
S.~Tsioutsiouliklis, E.~Pitoura, P.~Tsaparas, I.~Kleftakis, and N.~Mamoulis, ``Fairness-aware pagerank,'' in \emph{Proceedings of the Web Conference 2021}, 2021, pp. 3815--3826.

\bibitem{espin2022inequality}
L.~Esp{\'\i}n-Noboa, C.~Wagner, M.~Strohmaier, and F.~Karimi, ``Inequality and inequity in network-based ranking and recommendation algorithms,'' \emph{Scientific reports}, vol.~12, no.~1, p. 2012, 2022.

\bibitem{pastor2004evolution}
R.~Pastor-Satorras and A.~Vespignani, \emph{Evolution and structure of the Internet: A statistical physics approach}.\hskip 1em plus 0.5em minus 0.4em\relax Cambridge University Press, 2004.

\bibitem{santamaria2018comparison}
L.~Santamar{\'\i}a and H.~Mihaljevi{\'c}, ``Comparison and benchmark of name-to-gender inference services,'' \emph{PeerJ Computer Science}, vol.~4, p. e156, 2018.

\bibitem{blondel2008fast}
V.~D. Blondel, J.-L. Guillaume, R.~Lambiotte, and E.~Lefebvre, ``Fast unfolding of communities in large networks,'' \emph{Journal of statistical mechanics: theory and experiment}, vol. 2008, no.~10, p. P10008, 2008.

\bibitem{tsioutsiouliklis2022link}
S.~Tsioutsiouliklis, E.~Pitoura, K.~Semertzidis, and P.~Tsaparas, ``Link recommendations for pagerank fairness,'' in \emph{Proceedings of the ACM Web Conference 2022}, 2022, pp. 3541--3551.

\bibitem{tsioutsiouliklis2020fairness}
S.~Tsioutsiouliklis, E.~Pitoura, P.~Tsaparas, I.~Kleftakis, and N.~Mamoulis, ``Fairness-aware link analysis,'' \emph{arXiv preprint arXiv:2005.14431}, 2020.

\bibitem{leskovec2009community}
J.~Leskovec, K.~J. Lang, A.~Dasgupta, and M.~W. Mahoney, ``Community structure in large networks: Natural cluster sizes and the absence of large well-defined clusters,'' \emph{Internet Mathematics}, vol.~6, no.~1, pp. 29--123, 2009.

\bibitem{takac2012data}
L.~Takac and M.~Zabovsky, ``Data analysis in public social networks,'' in \emph{International scientific conference and international workshop present day trends of innovations}, vol.~1, no.~6, 2012.

\bibitem{saad1986gmres}
Y.~Saad and M.~H. Schultz, ``Gmres: A generalized minimal residual algorithm for solving nonsymmetric linear systems,'' \emph{SIAM Journal on scientific and statistical computing}, vol.~7, no.~3, pp. 856--869, 1986.

\bibitem{rozemberczki2019multiscale}
B.~Rozemberczki, C.~Allen, and R.~Sarkar, ``Multi-scale attributed node embedding,'' 2019.

\end{thebibliography}

\end{document}